\documentclass[12pt,a4paper]{article}
\usepackage{amsmath,amssymb,mathrsfs,mathtools,framed,esint}
\usepackage[colorlinks]{hyperref}
\usepackage{color}
\usepackage[all]{xy}

\newcommand{\comment}[1]{\vspace{5mm}\par
\framebox{\begin{minipage}[c]{.95 \textwidth} \tt\bfi #1
\end{minipage}}\vspace{5 mm}\par}
\newcommand{\rem}[1]{}

\newcommand{\de}{{\rm d}}

\newcommand{\z}{{\mathbf{z}}}

\newcommand{\bM}{{\mathbf{M}}}
\newcommand{\bX}{{\mathbf{X}}}
\newcommand{\bx}{{\mathbf{x}}}
\newcommand{\bv}{{\mathbf{v}}}

\newcommand{\bb}{{\boldsymbol{b}}}
\newcommand{\bA}{{\mathbf{A}}}

\newcommand{\bE}{{\mathbf{E}}}
\newcommand{\bB}{{\mathbf{B}}}
\newcommand{\bJ}{{\mathbf{J}}}

\newcommand{\bU}{ {\boldsymbol{U}} }
\newcommand{\bcalX}{ {\boldsymbol{\mathcal{X}}} }

\newcommand{\brho}{{\boldsymbol{\rho}}}

\newcommand{\bP}{{\mathbf{P}}}

\newcommand{\bz}{{\mathbf{z}}}

\newcommand{\bXi}{{\boldsymbol{\Xi}}}
\newcommand{\bxi}{{\boldsymbol{\xi}}}

\newcommand{\bfi}{\bfseries\itshape}

\newcommand{\beq}{\begin{equation}}
\newcommand{\eeq}{\end{equation}}
\newcommand{\ben}{\begin{eqnarray}}
\newcommand{\een}{\end{eqnarray}}

\begin{document}

\title{Variational approach to low-frequency kinetic-MHD\\in the current-coupling scheme}

\author{Joshua W. Burby$^1$, Cesare Tronci$^2$\\
\it\footnotesize $^1$Courant Institute of Mathematical Sciences, New York University, New York 10012, United States\\
\it\footnotesize $^2$Department of Mathematics, University of Surrey, Guildford GU2 7XH, United Kingdom
}

\date{}

\maketitle

\begin{abstract}  
Hybrid kinetic-MHD models describe the interaction of an MHD bulk fluid with an ensemble of hot particles, which obeys a kinetic equation. 
In this work we apply Hamilton's variational principle to formulate new current-coupling kinetic-MHD models in the low-frequency approximation (i.e. large Larmor frequency limit). More particularly, we formulate current-coupling  schemes, in which energetic particle dynamics are expressed in either guiding center or gyrocenter coordinates. When guiding center theory is used to model the hot particles, we show how energy conservation requires corrections to the standard magnetization term. On the other hand, charge and momentum conservation in gyrokinetic-MHD lead to extra terms in the usual definition of the hot current density as well modifications to conventional gyrocenter dynamics. All these new features arise naturally from the underlying variational structure of the proposed models.
\end{abstract}

\bigskip

{\footnotesize
\tableofcontents
}

\bigskip
\section{Nonlinear kinetic-MHD models}

The effects of high-temperature particles on MHD stability have been widely studied over several decades (see e.g. \cite{Chen,Cheng2,Coppi}) due to their practical importance in fusion studies. The formulation of a self-consistent  coupling between MHD and hot particle kinetics has always attracted much attention, since the early works on the linear regime. {\color{black}The hybrid kinetic-fluid description emerges as convenient since a fully kinetic treatment of all species would significantly increase the computational costs, while on the other hand a fluid closure for the energetic particles would totally neglect thermal effects and wave-particle interaction.} 
{\color{black}Over the years, two main kinetic-fluid coupling strategies} emerged naturally: pressure coupling \cite{Chen} and current coupling \cite{Coppi}. These two variants appeared to be interchangeable as long as linear models were being considered \cite{ParkEtAl}. However, the {\color{black}nonlinear} pressure coupling was further consolidated in the community and its mathematical footing was unfolded  \cite{Bri94} in terms of its underlying variational structure. 

The first nonlinear extensions of hybrid kinetic-MHD models appeared in the early 90's \cite{Cheng,ParkEtAl} and their consistency has been recently challenged in \cite{HoTr2011,Tronci2010,TrTaCaMo}. More particularly, upon adopting the assumption of full-orbit Vlasov trajectories for the hot particles, it was shown that the current-coupling scheme (CCS) of Vlasov-MHD is both mathematically and physically consistent, while the pressure-coupling scheme (PCS, most widely used in computer simulations) suffers from lacking an exact energy balance \cite{TrTaCaMo}. As a result, the PCS leads to unphysical instabilities.
New energy-conserving variants of the pressure-coupling scheme of Vlasov-MHD have been proposed \cite{HoTr2011,Tronci2010} and the unphysical instabilities appear to be removed in these variants \cite{TrTaCaMo}.

As mentioned above, the problems related to the nonlinear implementation of pressure-coupling simulations are removed in the current-coupling scheme of Vlasov-MHD. For example, a resistive {\color{black}CCS} variant is successfully implemented numerically by Belova {\color{black}\it et al.} \cite{BeGoFrTrCr,BelovaPark} and its mathematical foundations were recently explored further in \cite{ChSuTr}. However, fusion studies often make use of the drift-kinetic and gyrokinetic approximations, which drastically affect the structure of the equations of motion. {\color{black}In this regard, Littlejohn's pioneering work \cite{Littlejohn1981,Littlejohn1982,Littlejohn}} showed how inserting the approximations on Hamilton's variational principle \cite{Littlejohn} is the simplest and safest way of obtaining approximated descriptions in nonlinear dynamics.  This  variational approach was also exploited  by Brizard \cite{Brizard_JPP} and Sugama \cite{Sugama}, who provided the variational setting of gyrokinetic theory. More recently, this route has  unfolded some rather unexpected analogies between guiding center motion and the dynamics of nematic molecules in liquid crystal dynamics \cite{Tronci2016}.

This paper {\color{black}focuses on fully collisionless hybrid models. This assumption may not be completely satisfied even for weakly collisional plasmas (e.g. see \cite{BinTre} for the effects of collisions in galactic dynamics). Indeed, resistive extensions of hybrid MHD are available in the literature \cite{BelovaPark}, although the inclusion of resitivity in Ohm's law may not be completely straightforward (see e.g. Section 1.2 in \cite{ChSuTr}). In addition, a complete collisional treatment would necessarily involve collisions between energetic particles.  For example, a conservative gyrokinetic collision operator was recently developed in \cite{BuBrQi}, while no approach is currently available to address this question in the context of hybrid models. Given these difficulties we shall neglect collisional effects entirely. More specifically, this paper  provides new {\color{black}conservative} variants of the CCS by combining the variational description of ideal MHD \cite{Newcomb} with  the variational approach to both guiding center and gyrocenter motion for the  energetic particles.}
We believe such new variants are necessary because previous drift-kinetic-MHD and gyrokinetic-MHD models in the literature \cite{BeDeCh,ToSaWaWaHo} fail to possess a variety of basic conservation laws. In particular, no previous model exactly conserves energy, momentum, and hot charge simultaneously. 

After reviewing the variational structure of the CCS for Vlasov-MHD in Section \ref{sec:intro2}, we shall move on to drift-kinetic-MHD (Section \ref{sec:DK-MHD}). In this case, we shall show how a moving-dipole correction to the magnetization \cite{Ka,Pfirsch,PfMo} is crucial in ensuring an exact energy balance \cite{BrTr}. In Section \ref{sec:GK-MHD}, we pass to the gyrokinetic description. In order to ensure our theory is consistent with charge conservation, we shall present a new gauge-invariant Lagrangian underlying gyrokinetic theory. We will then combine our new gyrokinetic Lagrangian with Newcomb's MHD variational principle to obtain a {\color{black}conservative} CCS for gyrokinetic-MHD. This model remedies the issues with momentum and hot charge conservation present in previous work.

\section{Vlasov-MHD  \label{sec:intro2}}

\subsection{Formulation of the Vlasov-MHD model\label{SEC:Form1}}

The current-coupling scheme for kinetic-MHD was first presented in \cite{ParkEtAl} and we shall begin our treatment by reviewing its explicit formulation, under the assumption that the hot particles obey full-orbit Vlasov kinetics. This case was also treated in \cite{HoTr2011,Tronci2010,TrTaCaMo}. The starting point is the following kinetic multi-fluid system:
\begin{align}\label{multi-fluid-momentum1}
&\rho_s\partial_t\bU_s+\rho_s\left(\bU_s\cdot\nabla\right)\bU_s
=
a_s\rho_s\left(\bE+\bU_s\times\bB\right)-\nabla\mathsf{p}_s
\\
&
\partial_t\rho_s+\nabla\cdot\left(\rho_s\bU_s\right)=0
\\\label{multi-fluid-Vlasov1}
&
\partial_ t F+\mathbf{v}\cdot\nabla F+a_h(\bE+\bv\times\bB)\cdot\nabla_{\!\mathbf{v}} F=0
\\
&
\varepsilon_0\partial_t\bE=\mu_0^{-1}\nabla\times\bB-\sum_s a_s\rho_s\bU_s-
{\bf J}_h
\label{AmpereGC1}
\\
&
\partial_t\bB=-\nabla\times\bE
\\
&
\varepsilon_0
\nabla\cdot\bE=\sum_s a_s\rho_s+q_h n_h
\,,\qquad
\nabla\cdot\bB=0
\label{GaussGC1}
\end{align}
Here, $s=i,e$ denotes the fluid species (ions or electrons) with charge-to-mass ratio $a_s\color{black}=q_s/m_s$, mass density $\rho_s$, and partial pressure ${\sf p}_s$; $F(\bx,\bv)$ is the kinetic  density on phase-space {\color{black}and $q_h$ denotes the energetic particle charge}.  With this notation, we write the hot particle density and current density as
\[
n_h=\int\!F\,\de^3 \bv
\,,\qquad
{\bf J}_{h}=q_h\int\!\mathbf{v}\,F\,\de^3 \bv
\,.
\]
In order to derive the hybrid CCS, we proceed by inserting the usual assumptions underlying MHD. First, we assume quasineutrality by formally letting $\varepsilon_0\to0$ in Amp\`ere's law \eqref{AmpereGC1} and the Gauss law  in \eqref{GaussGC1}. Then, we  neglect the electron inertia by taking the limit $m_e\to 0$ for the second species ($s=e$) in the fluid  equation \eqref{multi-fluid-momentum1}. {\color{black}(See \cite{Freidbergbook} for alternative approaches in deriving the MHD limit).} Under this assumption, the sum of the equations \eqref{multi-fluid-momentum1} for $s=i,e$ yields
\begin{equation}\label{TotMom1}
\rho_i\partial_t\bU_i+\rho_i\left(\bU_i\cdot\nabla\right)\bU_i
=
\left(a_i\rho_i+a_e\rho_e\right)\bE+\left(a_i\rho_i\bU_i+a_e\rho_e\bU_e\right)\times\bB-\nabla(\mathsf{p}_i+\mathsf{p}_e)
\,,
\end{equation}
which then becomes (upon dropping the subscript $i$ {\color{black}and denoting $\bJ=\mu_0^{-1}\nabla\times\bB$})
\[
\rho\left(\partial_t\bU+\bU\cdot\nabla\bU\right)=-q_hn_h\bE+(\bJ-\bJ_h)\times\bB-\nabla(\mathsf{p}_i+\mathsf{p}_e)
\,.
\]
At this stage, in order to complete the formulation of the CCS, we need to express the electric field $\bE=-\partial_t\bA-\nabla\varphi$ by using Ohm's law \cite{Tronci2010,TrTaCaMo}. For the present purpose, we shall use ideal Ohm's law $\bE=-\bU\times\bB$ thereby obtaining the CCS momentum equation (denote ${\sf p}={\sf p}_i+{\sf p}_e$)
\[
\rho\left(\partial_t\bU+\bU\cdot\nabla\bU\right)=\Big(\bJ+q_hn_h\bU-\bJ_h\Big)\times\bB-\nabla{\sf p}
\,,
\]
along with the frozen-in condition $\partial_t\bB=\nabla\times(\bU\times\bB)$. {\color{black}We remark that the validity of ideal Ohm's law in the CCS is not entirely obvious; although we have introduced the usual approximations associated with ideal MHD, the presence of a hot kinetic species complicates the usual derivation of ideal Ohm's law from the electron momentum equation. However, the physical consistency of the present approach was shown} in Section 2.1 of \cite{TrTaCaMo},  in line with current efforts in hybrid MHD schemes. Thus, we shall adopt ideal Ohm's law throughout this work.
Eventually, one is left with the following set of equations for the current-coupling scheme of Vlasov-MHD:
\begin{align}\label{cc-hybrid-momentum}
&\rho\partial_t\bU+\rho\left(\bU\cdot\nabla\right)\bU = \left(q_hn_h\bU -\bJ_h+\mathbf{J}\right)\times\bB
-\nabla\mathsf{p}
\\
& \partial_t\rho+\nabla\cdot\left(\rho\bU\right)=0
\\
& \partial_t F+\bv\cdot\nabla F+a_h\left(\bv-\bU\right)\times\bB\cdot\nabla_{\!\bv} F=0 \label{cc-hybrid-Vlasov}
\\
& \partial_t\bB=\nabla\times\left(\bU\times\bB\right) \label{cc-hybrid-end} \,.
\end{align}
A resistive version of these equations are currently simulated in the ``{\it HY}brid and {\it M}HD simulation code'' (HYM) by Belova {\it\color{black} et al.} \cite{BeGoFrTrCr,BelovaPark}. Upon including resistive and viscous effects, the existence of weak solutions of this system has recently been established in \cite{ChSuTr}.

\subsection{The variational framework for Vlasov-MHD}
The variational theory of equations \eqref{cc-hybrid-momentum}-\eqref{cc-hybrid-end} was first presented in \cite{HoTr2011} by combining the variational structures of ideal MHD \cite{HoMaRa1998} and Vlasov kinetic theory \cite{Low,CeMaHo}. In recent years, the variational structure of Vlasov kinetics has been slightly modified \cite{SqQiTa,Tronci2014} so that it fits in the general theory of phase-space Lagrangians \cite{Littlejohn}: a review of the variational framework for collisionless kinetic theories is presented in the Appendix. In this framework, the Eulerian variational principle underlying the current-coupling Vlasov-MHD model reads
\beq\label{VarPrinc}
\delta\int_{t_1}^{t_2}\!\!\left(L_{\text{p}}+L_\textit{MHD}\right)\de t=0\,,
\eeq
where the Eulerian particle Lagrangian and MHD Lagrangian are given by
\begin{align}
L_{\text{p}}&=\int\!F(\z,t)\left[(m_h\bv+q_h\bA(\bx,t))\cdot\mathbf{u}(\z,t)-\frac{m_h}2|\bv|^2-q_h\varphi(\bx,t)\right]\,\de^6\mathbf{z}\\
L_\textit{MHD}&=\frac{1}2\int\!\rho\,|\bU|^{2\,}\de^3 \mathbf{x}-\int\!\rho\,\mathcal{U}(\rho)\,\de^3 \mathbf{x}-\frac1{2\mu_0}\int\!{|\nabla\times \bA|^2}\,\de^3\mathbf{x}.
\label{MHD-Lagrangian}
\end{align}
Here, the notation is as follows: $\z=(\bx,\bv)$ denotes the Eulerian coordinate in phase space, $F(\z,t)$ is the Eulerian Vlasov density (we use capital $F$ for later convenience), and $\varphi(\bx,t)$ is the electrostatic potential expressed in the \emph{hydrodynamic gauge}
\beq\label{hygauge}
\varphi=\bU\cdot\bA
\,,
\eeq
occurring in plasma fluid models \cite{IlLa,Tronci2014,Davignon2016}.  {\color{black}In the present work, this gauge is exploited in combination with the frozen-in vector potential for consistency with ideal Ohm's law.} As in the previous section, $\bU(\bx,t)$ represents the Eulerian bulk velocity, $\rho(\bx,t)$ is the mass density of the bulk fluid, and $\mathcal{U}$ is the internal MHD energy. {\color{black}Notice that here we assume a barotropic bulk flow so that the internal energy is a function only of the density, i.e. $\mathcal{U}=\mathcal{U}(\rho)$; this simplifying assumption can be easily relaxed to allow for adiabatic flows including entropy dynamics by following the same procedure outlined in \cite{HoMaRa1998}}. In terms of the Lagrangian particle path in phase space $\boldsymbol{z}(\z_0,t)$ and the Lagrangian path of the MHD bulk $\boldsymbol{q}(\bx_0,t)$, one has respectively two Eulerian vector fields $\boldsymbol{\mathcal{X}}$ and $\boldsymbol{U}$ such that
\begin{align}
\partial_t\boldsymbol{z}(\z_0,t)&=\boldsymbol{\mathcal{X}}(\boldsymbol{z}(\z_0,t),t)\label{phase_space_velocity_def}\\
\partial_t\boldsymbol{q}(\bx_0,t)&=\boldsymbol{U}(\boldsymbol{q}(\bx_0,t),t),\label{configuration_space_velocity_def}
\end{align} 
where  $\bcalX(\z,t)$ has components
\begin{align}\label{EulXfield}
\boldsymbol{\mathcal{X}}(\z,t)=\Big(\mathbf{u}(\z,t),\,\mathbf{a}(\z,t)\Big)\,.
\end{align}
{\color{black}The relation between Eulerian and Lagrangian coordinates is at the heart of Euler-Poincar\'e theory \cite{HoMaRa1998}; for further details on the use of Lagrangian path coordinates in plasma physics, we address the reader to \cite{Morrison2005,Morrison2}.}

The Eulerian variational structure of \eqref{VarPrinc} arises from its corresponding Lagrangian formulation \cite{HoTr2011} by exploiting the relabeling symmetry of both the MHD and the phase-space Vlasov actions in \eqref{VarPrinc}. This reduction by symmetry is  a standard tool in continuum theories and it is known as Euler-Poincar\'e reduction \cite{HoMaRa1998}. This process enforces the following variational relations
\beq\label{variations-A}
\delta\bU=\partial_t\boldsymbol{\xi}+(\bU\cdot\nabla)\boldsymbol{\xi}-(\boldsymbol{\xi}\cdot\nabla)\bU
\,,\qquad\ 
\delta\boldsymbol{\mathcal{X}}=\partial_t\boldsymbol\Xi+(\boldsymbol{\mathcal{X}}\cdot\nabla_\z)\boldsymbol{\Xi}-(\boldsymbol\Xi\cdot\nabla_\z)\boldsymbol{\mathcal{X}}
\eeq
where $\boldsymbol\Xi(\z,t)$ and  $\boldsymbol{\xi}(\bx,t)$ are arbitrary Eulerian vector fields such that $\delta{\boldsymbol{z}}(\z_0,t)=\boldsymbol\Xi(\boldsymbol{z}(\z_0,t),t)$ and $\delta \boldsymbol{q}(\bx_0,t)=\boldsymbol{\xi}(\boldsymbol{q}(\bx_0,t),t)$. 
These relations are accompanied by the constrained variations
\beq
\delta \rho=-\nabla\cdot(\rho\boldsymbol{\xi})
\,,\qquad
\delta F=-\nabla_\z\cdot(F\boldsymbol\Xi)
\,,\qquad
\delta \bA =\boldsymbol{\xi}\times\nabla\times\bA-\nabla(\boldsymbol{\xi}\cdot\bA)
\,.
\label{variations1}
\eeq
The expressions above arise naturally by taking variations of the following Lagrangian relations
\beq\label{pullbacks}
\rho(\boldsymbol{q},t)\,\de^3 \boldsymbol{q} = \rho_0(\bx_0)\,\de^3 \mathbf{x}_0
\,,\qquad\ 
F(\boldsymbol{z},t)\,\de^6 \boldsymbol{z} = F_0(\z_0)\,\de^6 \mathbf{z}_0
\,,\qquad\ 
\bA(\boldsymbol{q},t)\cdot\de\boldsymbol{q}=\bA_0(\bx_0)\cdot\de\bx_0
\,,
\eeq
where $\boldsymbol{q}=\boldsymbol{q}(\bx_0,t)$ and $\boldsymbol{z}=\boldsymbol{z}(\z_0,t)$. 
Then, upon proceeding in exact analogy, one can also take the time derivative of \eqref{pullbacks} to obtain the advection equations
\beq
\partial_t \rho=-\nabla\cdot(\rho\boldsymbol{U})
\,,\qquad
\partial_t F=-\nabla_\z\cdot(F\boldsymbol{\mathcal{X}})
\,,\qquad
\partial_t \bA = \boldsymbol{U}\times\nabla\times\bA-\nabla(\boldsymbol{U}\cdot\bA)
\,,
\label{advection1}
\eeq
which are to be coupled to the equations for $\bU(\bx,t)$ and $\boldsymbol{\mathcal{X}}(\bz,t)$. The latter are obtained by using the variations \eqref{variations-A} in the variational principle \eqref{VarPrinc}. {\color{black}We remark that here we are following the standard Euler-Poincar\'e procedure as it is described in detail in Section 7 of \cite{HoMaRa1998} (see equation (7.2) therein).}

By  setting the first variation of the action equal to zero {\color{black}as in \eqref{VarPrinc}}, one obtains  \cite{HoTr2011}

\begin{align}\label{EP-CCS1}
&\rho\left(\partial_t+\bU\cdot\nabla\right)\left(\bU-q_hn_h\rho^{-1}\bA\right)+q_hn_h\nabla\bU\cdot\bA
=-\nabla\mathsf{p}
\nonumber
\\
&\hspace{6.8cm}
-(\bJ_h-q_hn_h\bU-\mathbf{J})\times\bB
+\bA\,\nabla\cdot(\bJ_h-q_hn_h\bU)
\,,
\\
\label{EP-CCS2}
&q_h\partial_t \bA+q_h\mathbf{u}\cdot\nabla\bA+\nabla\mathbf{u}\cdot(m_h\bv+q_h\bA)+m_h\mathbf{a}-\nabla(m_h{\bf u}\cdot\bv+q_h{\bf u}\cdot\bA-q_h\bU\cdot\bA\big)=0
\,,
\\
\label{EP-CCS2-bis}
&\mathbf{u}=\bv
\end{align}
where we have introduced the definitions
\[
{\sf p}=\rho^2\mathcal{U}'(\rho)
\,,\qquad\quad
\bB=\nabla\times\bA
\,,\qquad\quad
\mathbf{J}=\mu_0^{-1}\nabla\times\bB
\,,
\]
and 
\beq\label{kinmom-def}
n_h=\int \!F\,\de^3 \mathbf{v}
\,,\qquad\quad
\bJ_h=\frac{\delta L_{\text{p}}}{\delta \bA}=q_h\int \!F\,\mathbf{u}\,\de^3\mathbf{v}.
\eeq
Here, we have used the functional derivative notation (see Appendix).  Now, replacing \eqref{EP-CCS2-bis} into \eqref{EP-CCS2} yields
\[
\mathbf{a}(\bz,t)=-a_h\left(\partial_t {\bf A}+\nabla(\bU\cdot\bA)\right)+a_h\bv\times\bB
\,.
\]
In conclusion, by the third in \eqref{advection1}, we obtain
\beq
\boldsymbol{\mathcal{X}}(\bz,t)=\Big(\bv,\,a_h(\bv-\bU)\times\bB\Big)
\,.
\label{mark}
\eeq
Also, by using the relation $\partial_t n_h=-q_h^{-1}\nabla\cdot\bJ_h$ (verified explicitly from the second of \eqref{advection1}) and
\beq\label{helpfulrel}
\partial_t\left(n_h\rho^{-1}\bA\right)=\bA\rho^{-1}\,\nabla\cdot(n_h\bU-q_h^{-1}\bJ_h)
\,,
\eeq
in \eqref{EP-CCS1}, {\color{black}
we obtain equation  \eqref{cc-hybrid-momentum}. Also, inserting \eqref{mark} in the second of \eqref{advection1} and noticing that $\nabla_{\bz}\cdot\boldsymbol{\cal X}=0$ yields equation \eqref{cc-hybrid-Vlasov}. Furthermore, taking the curl of the third in \eqref{advection1} returns the advection law \eqref{cc-hybrid-end}. Thus, we have obtained the final equations \eqref{cc-hybrid-momentum}-\eqref{cc-hybrid-end}  of the hybrid Vlasov-MHD model in the current-coupling scheme \cite{HoTr2011,ParkEtAl,Tronci2010}. Notice that by using Euler-Poincar\'e reduction we were able to write our new model in terms of the purely Eulerian variables  $(\bU,\boldsymbol{\cal X},\rho,\bA,F)$, as opposed to their Lagrangian counterparts $(\boldsymbol{q},\dot{\boldsymbol{q}},\boldsymbol{z},\dot{\boldsymbol{z}},\rho_0,\bA_0,F_0)$. Specifically, the fixed parameters  $(\rho_0,\bA_0,F_0)$ generated their corresponding Eulerian advected quantities by the relations \eqref{pullbacks}.}

In the low frequency limit, one introduces guiding center coordinates or gyrocenter coordinates on the reduced phase space \cite{BrHa,CaBr}, while the dynamics is parameterized by the magnetic moment magnitude $\mu$.
The above treatment is adapted to this case without essential modifications. Indeed, the Euler-Poincar\'e theory of the Maxwell-Vlasov system in the guiding center approximation was recently presented explicitly in \cite{BrTr}, while the correspondent formulation of gyrokinetics appeared in \cite{SqQiTa}. 
These cases differ  by the way the analogues of the canonical momentum $m_h\bv+q_h\bA$ and the kinetic energy $m_h|\bv|^2/2$ are written explicitly. 
\rem{ 
Then, the whole construction is analogous to the one above upon replacing the six-dimensional Eulerian coordinate $\z=(\bx,\bv)$ by $\z_{\rm gc}=({\bf X},u_\parallel)$, so that $\boldsymbol{\mathcal{X}}(\z,t)$ will be replaced by $\boldsymbol{\mathcal{X}}_\textrm{gc}(\z_\textrm{gc},t;\mu)$ (here, we have emphasized the parametric dependence on the magnetic moment magnitude $\mu$). However, a special remark on the canonical momentum $\boldsymbol{P}(\z,\bA)$ is necessary at this point. Indeed, 
} 
 This point will be an important difference, which on the other hand can be dealt with using standard differentiation rules.

\section{Drift-kinetic-MHD\label{sec:DK-MHD}}
The simplest low-frequency approximation of the Vlasov equation is provided by guiding center theory \cite{Littlejohn}.
In this section, we shall present a new current-coupling scheme that couples MHD to a drift kinetic equation for guiding center trajectories. Our formulation of the drift kinetic equation will be based on a variant of guiding center theory that assumes all components of the electric field are asymptotically small; for guiding center theory that allows the $E\times B$ speed to be comparable to the thermal velocity, see \cite{Littlejohn1981}. For completeness, we shall present two different formulations of the model: the first will be constructed by working on the equations of motion, while the second will be based on the variational approach. 

\subsection{Formulation of the new drift-kinetic-MHD model}
In order to derive the CCS in the guiding center approximation, we shall start from first principles by coupling ordinary fluid equations to the drift kinetic equation for guiding center orbits and to Maxwell's equations. More explicitly, we shall start with the kinetic multi-fluid system \eqref{multi-fluid-momentum1}-\eqref{GaussGC1} by replacing the Vlasov equation \eqref{multi-fluid-Vlasov1} by its guiding center approximation. In turn, a guiding center ensemble behaves as a magnetized medium (this concept was also developed further in  \cite{Morrison_gv_2014} for gyroviscous MHD). This implies that the modified expression ${\bf J}_h={\bf J}_{\rm gc}+\nabla\times{\bf M}_{\rm gc}$ for the hot current is given by the sum of the current carried by the guiding centers and their intrinsic magnetization {\color{black}contribution. The latter is due to the particle gyration and is usually given as $-\nabla\times\int_{\mu\,}\mu\bb\,F\,\de v_\|$ (in standard guiding center notation), although we shall see how this expression requires appropriate corrections}. Notice that if we used the variant of guiding center theory allowing the $E\times B$ speed to be comparable to the  thermal velocity of the guiding center ensemble, then the magnetized medium would also acquire an electric polarization so that the hot current would read  ${\bf J}_h={\bf J}_{\rm gc}+\nabla\times{\bf M}_{\rm gc}+\partial_t\bP_h$  (and a polarization correction would also appear in the definition of the total  charge). 
While polarization effects will be crucial in Section \ref{sec:GK-MHD} for the development of gyrokinetic-MHD, for the moment we shall follow Littlejohn's original approach \cite{Littlejohn} so that $\bP_h=0$. We emphasize that this Section adopts the standard guiding center assumption that a typical gyroradius is much smaller than the magnetic field scale length, i.e. $\rho_L\ll B/|\nabla\bB|$.

The equations \eqref{multi-fluid-Vlasov1}-\eqref{AmpereGC1} in the kinetic multi-fluid system \eqref{multi-fluid-momentum1}-\eqref{GaussGC1} are then modified as follows:
\begin{align}
\label{multi-fluid-Vlasov}
&
\partial_t F+\nabla\cdot(F{\bf u}_{\rm gc})+\partial_{v_\parallel} (F{a}_{\parallel\text{gc}})=0
\\
&
\varepsilon_0\partial_t\bE=\mu_0^{-1}\nabla\times\bB-\sum_s a_s\rho_s\bU_s-
{\bf J}_{\rm gc}-\nabla\times{\bf M}_{\rm gc}
\label{AmpereGC}
\end{align}
Here,  $F(\bX,v_\|;\mu)$ is the kinetic  density on the guiding center phase-space and ${\bf u}_{\rm gc}(\bX,v_\|;\mu)$ is the Eulerian guiding center velocity, so that  $\bb\cdot{\bf u}_{\rm gc}=v_\|$ and ${\bf u}_{\perp\rm gc}=-\bb\times\bb\times{\bf u}_{\rm gc}$ are its parallel and perpendicular components, respectively (recall the standard guiding center notation $\bb=\bB/B$). Also, $a_{\parallel\text{gc}}(\bX,v_\|;\mu)$ is the Eulerian parallel acceleration. 
Explicitly, the components of the Eulerian phase-space vector field read
\beq\label{EulerianVF}
{\bf u}_{\rm gc}(\bX,v_\|;\mu)=\frac1 {B^*_\|}\Big(v_\|\bB^*-\bb\times\bE^*\Big)
\,,\qquad \ 
a_{\|\rm gc}(\bX,v_\|;\mu)= \frac{a_h}{B^*_\|}\,\bB^*\cdot\bE^*,
\eeq
where
\beq
\bB^*:=\nabla\times\left(\bA+a_h^{-1}v_\parallel \bb\right)
\,,\qquad\ 
\bE^*:=-\partial_t\left(\bA+a_h^{-1}v_\parallel \bb\right)-\nabla(\varphi+\mu B)
\label{EffFields}
\eeq
and all field variables in \eqref{EulerianVF} are evaluated at the guiding center position $\bX$. Notice that the above relations are the Eulerian correspondent of the equations \eqref{SP-EQ} in Appendix \ref{sec:appGC} for a single guiding center trajectory.

 We recall that all guiding center quantities depend  on the magnetic moment invariant $\mu$ and we insert the notation 
\[
\int_{\mu}\mathcal{F}(\bX,v_\|;\mu)\,\de v_\|:=\iint\!\de\mu\,\de v_\|\,\mathcal{F}(\bX,v_\|;\mu)
\] 
for any function $\mathcal{F}=\mathcal{F}(\bX,v_\|;\mu)$. With this notation, we write the guiding center density and current as
\[
n_h =\int_{\!\mu\,} \!F\,\de v_\parallel
\,,\qquad
{\bf J}_{\rm gc}=q_h\int_{\!\mu\,}\!\mathbf{u}_{\rm gc}\,F\,\de v_\|
\,,
\]
while the explicit expression of the guiding center magnetization ${\bf M}_{\rm gc}$ will be given later on.  
\rem{ 
Notice that the last magnetization term in Amp\`ere's law \eqref{AmpereGC} is necessary to ensure energy and momentum conservation in the presence of self-evolving electromagnetic fields. This moving dipole correction to the usual magnetization vector ${\bf M}=-\int_{\mu\,}\mu\bb\,F\delta_\bx\,\de v_\|$ has been known for decades \cite{Ka,Pfirsch,PfMo} and its fundamental role was recently emphasized in \cite{BrTr}.
}  

In order to derive the hybrid CCS, we follow the same steps illustrated in Section \ref{SEC:Form1}. First, we assume quasineutrality by letting $\varepsilon_0\to0$ in  \eqref{AmpereGC} and \eqref{GaussGC1}. Then, we  take the limit $m_e\to 0$ {\color{black}(equivalent to neglecting electron inertia)}  in the fluid  equation \eqref{multi-fluid-momentum1} for $s=e$. Under this assumption, the sum of the equations \eqref{multi-fluid-momentum1}  yields \eqref{TotMom1}, which then becomes 
\[
\rho\left(\partial_t\bU+\bU\cdot\nabla\bU\right)=-q_hn_h\bE+(\bJ-\bJ_{\rm gc}-\nabla\times{\bf M}_{\rm gc})\times\bB-\nabla(\mathsf{p}_i+\mathsf{p}_e)
\,.
\]
At this stage, we use Ohm's law  $\bE=-\bU\times\bB$ thereby obtaining the CCS momentum equation 
\beq
\rho\left(\partial_t\bU+\bU\cdot\nabla\bU\right)=\Big(\bJ+q_hn_h\bU-\bJ_{\rm gc}-\nabla\times{\bf M}_{\rm gc}\Big)\times\bB-\nabla{\sf p}
\,,
\eeq
along with the frozen-in condition \eqref{cc-hybrid-end}. 
Notice that ideal Ohm's law leads to the effective electric field
\beq
\bE^*=-\bU\times\bB-\frac{v_\|}{a_h B}\,\bb\times\bb\times\nabla\times(\bU\times\bB)-\mu\nabla B
\,,
\label{effE}
\eeq
which must be replaced in \eqref{EulerianVF} in order to solve for the guiding center motion in the drift-kinetic equation \eqref{multi-fluid-Vlasov}. Eventually, one is left with the following set of equations for the CCS in the guiding center approximation:
\begin{align}\label{cc-hybrid-momentum-GC}
&\rho\partial_t\bU+\rho\left(\bU\cdot\nabla\right)\bU
=\Big(\bJ+q_hn_h\bU-\bJ_{\rm gc}-\nabla\times{\bf M}_{\rm gc}\Big)\times\bB-\nabla{\sf p}
\\
& \partial_t\rho+\nabla\cdot\left(\rho\bU\right)=0 \label{cc-hybrid-mass}
\\
&\label{cc-hybrid-mass-DK}
\partial_t f+\frac1 {B^*_\|}\Big(v_\|\bB^*-\bb\times\bE^*\Big)\cdot\nabla f+\frac{a_h}{B^*_\|}\left(\bB^*\cdot\bE^*\right)\partial_{v_\parallel} f=0
\\
&
\partial_t\bB=\nabla\times\left(\bU\times\bB\right)
\label{cc-hybrid-end-GC}
\,,
\end{align}
where we introduced the reduced distribution function $f=F/B^*_\|$, so that
\beq
\bJ_{\rm gc}=q_h\int_{\!\mu\,}\!\Big(v_\|\bB^*-\bb\times\bE^*\Big)f\,\de v_\|
\label{J+M}
\eeq
and we recall \eqref{effE}. Here, we have used Liouville's theorem in the form \cite{CaBr}
\begin{align}
\partial_t B^*_\|+\nabla\cdot(B^*_\|{\bf u}_{\rm gc})+\partial_{v_\parallel}(B^*_\|a_{\parallel\text{gc}})=0
\,.\label{liouville_gc}
\end{align}
Notice that \eqref{cc-hybrid-mass-DK} is the kinetic equation for an ensemble of guiding center trajectories: see equations \eqref{SP-EQ} in Appendix \ref{sec:appGC}.

At this point, in order to complete the system, one needs to specify the explicit form of the guiding center magnetization ${\bf M}_{\rm gc}$. 
In the standard theory of guiding center motion, as it was derived in \cite{Littlejohn}, the magnetic field is external and the magnetization vector is given by the total magnetization $-\int_\mu \mu\,\bb \,F\,\de v_\|$ \cite{CaBr}. This is precisely the expression that has been used in the hybrid simulations by Todo {\it\color{black} et al.} \cite{Todo}-\cite{ToVZBiHe}. However, when guiding center motion is coupled to Maxwell's equations for the self-consistent evolution of the electromagnetic field, the magnetization carries a moving-dipole correction, which ensures both energy and momentum conservation. Explicitly, the guiding center magnetization reads
\begin{align}\nonumber
{\bf M}_{\rm gc}&=-\int_{\!\mu\,}\!\Big(\mu\bb-\frac{m_hv_\|}{B}\mathbf{u}_{\rm gc}^\perp\Big)F\,\de v_\|
\\
&=-\int_{\!\mu\,}\!\left[\mu B^*_\|\bb-\frac{m_hv_\|}{B}\Big(v_\|\bB^*_\perp-\bb\times\bE^*\Big)\right]f\,\de v_\|
\label{magn-corr}
\end{align}
where  $\bB^*_\perp=-(m_h/q_h)v_\|\,\bb\times\bb\times\nabla\times\bb$. We remark that the moving dipole correction to the usual magnetization vector $-\int_{\mu\,}\mu\bb\,F\,\de v_\|$ has been known for decades \cite{Ka,Pfirsch,PfMo} and its fundamental role was recently emphasized in \cite{BrTr}.

 The equations \eqref{cc-hybrid-momentum-GC}-\eqref{cc-hybrid-end-GC}, together with the relations \eqref{effE}, \eqref{J+M} and \eqref{magn-corr}, form a new current-coupling scheme for hybrid kinetic-MHD in the guiding center approximation for hot particle orbits. Notice that this system preserves the family of invariants $\int \!\Phi(f) B_\parallel^*\, dv_\parallel\,d^3\bX$ (where $\Phi(f)$ is an arbitrary function of $f$; e.g. $\Phi(f)=f\log f$ returns entropy conservation)
\rem{ 
\comment{CT: is this really what we mean? Shouldn't the invariants be of the type $\int \!\Phi(F) \, dv_\parallel\,d^3\bX$? Aren't these the Casimirs for the standard gc-Vlasov bracket? (say for external magnetic fields).\\
JB: The short answers: Yes. No. Yes, but both expressions are the same for full orbit (provided we use $(x,v)$-coordinates in full orbit).\\
The proof: Because $f$ is advected by $\mathcal{X}$ as a scalar, $\Phi(f)$ is also advected by $\mathcal{X}$ as a scalar. Therefore 
\begin{align*}
\frac{d}{dt}\int \Phi(f) B_\parallel^* dv_\parallel d^3\bX&=-\int \big(L_{\mathcal{X}}\Phi(f)\big)\,B_\parallel^*dv_\parallel d^3\bX +\int \Phi(f)\,\partial_t B_\parallel^* dv_\parallel d^3\bX\\
&=\int \Phi(f)\,\Big(\text{div}(B_\parallel^*\mathcal{X})+\partial_tB_\parallel^*\Big)dv_\parallel d^3\bX\\
&=0,
\end{align*}
where Eq.\,(\ref{liouville_gc}) was used in the last line. 
}\noindent
} 
in addition to the standard expressions for both cross helicity and magnetic helicity, as may be verified by a direct calculation. This is perhaps not surprising since these features are already present in the case of Vlasov kinetics for the hot particles, as shown in \cite{HoTr2011}. We also remark that the equations \eqref{cc-hybrid-momentum-GC}-\eqref{cc-hybrid-end-GC} differ from analogous models previously appeared in the literature \cite{PaBeFuTaStSu,Todo} by three main features:
\begin{itemize}

\item {\bf Standard guiding center theory:} unlike the treatment in \cite{PaBeFuTaStSu,ToSa,ToSaWaWaHo}, the parallel component of the effective magnetic field $B^*_\|$ is nowhere approximated by $B$ {\color{black}in order to retain energy conservation}. This is consistent with the CCS in \cite{Todo} and it follows from Littlejohn's standard theory of guiding center motion \cite{Littlejohn}.

\item {\bf $\bf E\boldsymbol{\times} B\boldsymbol{-}$drift current:} as a consequence of the previous item, the Lorentz force term $q_hn_h\bU\times\bB$ is retained in the fluid equation \eqref{cc-hybrid-momentum-GC}. Had we assumed  $B^*_\|\simeq B$, this Lorentz force term would cancel with the $E\times B-$drift current contribution $\bJ_\textit{\tiny \!E$\times\!$B}$ in \eqref{cc-hybrid-momentum-GC}, since \cite{ToSa,ToSaWaWaHo}
\begin{align}\nonumber
\left(q_hn_h\bU-\bJ_\textit{\tiny \!E$\times\!$B}\right)\times\bB=&
-q_h\bB\times\int_{\!\mu\,}\!\bigg(\bU +\frac{1}{B^*_\|}\bb\times\bE\bigg)F \de v_\|
\\
=&\  
q_h\left[\int_{\!\mu\,}\!\bigg(1-\frac{B}{B^*_\|}\bigg)\, F \de v_\|\right]\bU\times\bB
\,.
\label{LforceTerms}
\end{align}
For example, even if it does not affect the energy balance, the above term should be retained in the CCS presented in \cite{Todo}, while it appears to have been overlooked over the decades \cite{ToBeBr,ToVZHe,ToVZBiHe}.

\item {\bf Energy conservation:} unlike previous approaches, the guiding center magnetization retains here the moving-dipole correction $m_h\int_{\!\mu\,}\!B^{-1}{v_\|}\mathbf{u}_{\perp\rm gc}\, F\,\de v_\|$ in \eqref{magn-corr} to ensure energy and momentum balance \cite{BrTr,Ka,Pfirsch,PfMo}. For example, one can verify that omitting the corresponding term in \eqref{cc-hybrid-momentum-GC} yields the  (sign-indefinite) energy rate 
\[
\dot{E}=-\int\!\bigg[\left(\bU\times\bB\right)\cdot\nabla\times\!\int_{\!\mu\,}\!B^*_\|\,\frac{m_hv_\|}{B}\,\mathbf{u}_{\perp\rm gc}\,f\,\de v_\|\bigg]\de^3\bx
\,,
\]
where 
\[
E=\frac12\int\!\rho|\bU|^2\,\de^3\bx+\iint_{\!\mu\,}\!B^*_{\|\!}\left(\frac{m_h}2v_\|^2+\mu B \right)f\,\de v_\|\,\de^3\bx+\int\!\rho\,\mathcal{U}(\rho)\,\de^3 \bx+\frac1{2\mu_0}\int\! |\bB|^2\,\de^3 \bx
\,.
\]
We remark that this result contradicts the corresponding result in \cite{Todo} and it is unaffected by the presence of the terms \eqref{LforceTerms} in \eqref{cc-hybrid-momentum-GC} and/or by the replacement of $B^*_\|$ by $B$.
\end{itemize}

The fact that the new CCS conserves both energy and momentum is naturally inherited by its underlying variational structure, which follows by similar arguments as those presented in the previous section. This is the subject of the discussion below.

\subsection{The variational framework for drift-kinetic-MHD}
This section presents the variational formulation of the CCS \eqref{cc-hybrid-momentum-GC}-\eqref{cc-hybrid-end-GC}. A fundamental approach would require starting with \eqref{VarPrinc} and applying the guiding center approximation on the terms in the square brackets. For example, this can be achieved by resorting to the Klimontovich approach so that Lie perturbation theory \cite{CaLi,Littlejohn1981,Littlejohn1982} can be applied to single-particle orbits. To lowest order, an alternative to Lie perturbation techniques can be found in Littlejohn's variational approach \cite{Littlejohn} or its newly established variant in \cite{Tronci2016}. In order to avoid unnecessary digressions into the mathematics of the guiding center approximation, here we shall simply show how the equations \eqref{cc-hybrid-momentum-GC}-\eqref{cc-hybrid-end-GC} arise naturally from the following Hamilton's principle of the type \eqref{VarPrinc}, where  the Eulerian particle Lagrangian is now given by
\begin{multline}\label{GCparticleL}
L_{\text{p}}=\iint_\mu \!F(\bz,t)\,\Big\{\!\left[m_hv_\|\bb(\bX,t)+q_h\bA(\bX,t)\right]\cdot\mathbf{u}_{\text{gc}}(\z,t)
\\
-\frac{m_h}2\,v_\|^2-\mu B(\bX,t)-q_h\varphi(\bX,t)\Big\}\,\de^4\mathbf{z}.
\end{multline}
{\color{black}We remind the reader that in writing the Lagrangian above we have assumed that the $E\times B$ speed is much smaller than the thermal speed. If we were to allow for these two speeds to be comparable, it would suffice to simply subtract the $E\times B$ energy term $m_h B^{-2}\left|\bE\times\bb\right|^{2}\!/2$ from the guiding center kinetic energy $m_hv_\|^2/2+\mu B$ (e.g., see equation (3.49) in \cite{CaBr}), thereby producing extra magnetization terms along with polarization effects \cite{Krommes2}.

In \eqref{GCparticleL}}, we  denoted the guiding center phase-space coordinates by
\[
\bz=(\bX,v_\parallel)
\]
and we  also assumed the hydrodynamic gauge \eqref{hygauge}. In addition,  $\mathbf{u}_{\rm gc}(\z,t)$ and $a_{\|\rm gc}(\z,t)$ are now combined into the Eulerian (phase-space) guiding center vector field
\[
\boldsymbol{\mathcal{X}}_{\rm gc}(\z,t)=\Big(\mathbf{u}_{\rm gc}(\z,t),\,{a}_{\parallel\text{gc}}(\z,t)\Big)\,,
\]
in analogy to \eqref{EulXfield}. As in Eq.\,(\ref{phase_space_velocity_def}), we have $\dot{\boldsymbol{z}}(\z_0,t)=\boldsymbol{\mathcal{X}}_{\rm gc}(\boldsymbol{z}(\z_0,t),t)$, where ${\boldsymbol{z}}(\z_0,t)=\big(\boldsymbol{\mathscr{X}}(\z_0,t),\mathcal{V}_\|(\z_0,t)\big)$ denotes the Lagrangian particle path in phase space and $\z_0=(\bX_{0},{v_{\|0}})$ is the particle phase-space label.
 Notice that Lagrangian paths are all parameterized by the magnetic moment invariant $\mu$, which then should appear as an extra label (although it has been omitted in the present notation). However, this does not produce essential difficulties or modifications in the treatment.
Indeed, at this point, the problem set up is completely analogous to that of section \ref{sec:intro2}, including the variational relations \eqref{variations-A}-\eqref{variations1} and the advection equations \eqref{advection1}. The second relations in  \eqref{variations-A} and \eqref{variations1} now involve an Eulerian displacement vector field ${\boldsymbol{\Xi}}_{\rm gc}$ on the guiding center phase-space, so that $\delta\boldsymbol{\mathcal{X}}_{\rm gc}=\partial_t{\boldsymbol\Xi}_{\rm gc}+(\boldsymbol{\mathcal{X}}_{\rm gc}\cdot\nabla_\z)\boldsymbol{\Xi}_{\rm gc}-({\boldsymbol\Xi}_{\rm gc}\cdot\nabla_\z){\boldsymbol{\mathcal{X}}}_{\rm gc}
$ and $\delta F=-\nabla_\z\cdot(F{\boldsymbol\Xi}_{\rm gc})
$. Then, the MHD equation \eqref{EP-CCS1} now involves the hot current 
\begin{align}\nonumber
\bJ_h&=\frac{\delta L_{\text{p}}}{\delta \bA}+\nabla\times \frac{\delta L_{\text{p}}}{\delta \bB}\\
&=q_h\int_{\!\mu\,} \!F\,{\bf u}_{\rm gc}\,\de v_\|
+
\nabla\times\int_{\!\mu\,} \!F\left(m_hv_\|\frac{\partial b_i}{\partial \bB}\,u_{\rm gc}^i-\mu\frac{\partial B}{\partial \bB}\right)\,\de v_\|,
\label{hotcurrgc}
\end{align}
in analogy to the second of \eqref{kinmom-def}. 
The first equality above follows from the fact that the guiding center Lagrangian \eqref{GCparticleL} depends on the vector potential $\bA$ as well as on its curl. As it was noticed in \cite{BrTr}, this last dependence is responsible for the moving-dipole correction in the guiding center magnetization ${\bf M}_{\rm gc}=\delta L_{\text{p}}/\delta\bB$. Then, a chain-rule calculation shows  that 
\[
\bJ_h=\bJ_{\rm gc}+\nabla\times{\bf M}_{\rm gc}
\,,
\]
with the definitions in \eqref{J+M} and \eqref{magn-corr}.

For guiding center dynamics, the equations \eqref{EP-CCS2}-\eqref{EP-CCS2-bis} are replaced by (see Appendix)
\begin{multline*}
m_hv_\parallel\partial_t \bb+q_h\partial_t \bA+{\bf u}_\textrm{gc}\cdot\nabla(m_hv_\parallel\bb+q_h\bA)+{a_{\parallel\rm gc}}\bb+\nabla{\bf u}_\textrm{gc}\cdot (m_hv_\parallel \bb+q_h\bA)
\\
-\,\nabla \big(m_hv_\|\bb\cdot{\bf u}_{\rm gc}+q_h\bA\cdot{\bf u}_{\rm gc}-q_h\mu B-q_h\bU\cdot\bA\big)=0.
\end{multline*}
as well as $\bb\cdot{\bf u}_\textrm{gc}=v_\parallel$. Upon recalling the third in \eqref{advection1}, we have
\begin{align*}
&m_hv_\parallel\partial_t \bb+q_h\bU\times\bB+{a_{\parallel\rm gc}}\bb-{\bf u}_\textrm{gc}\times(q_h\bB+m_hv_\parallel\nabla\times\bb)+q_h\mu\nabla B=0\,,
\end{align*}
and since
\[
\partial_t \bb=-\frac{1}{B}\bb\times\bb\times\nabla\times\partial_t\bA
=-\frac{1}{B}\bb\times\bb\times\nabla\times(\bU\times\bB)
\,,
\]
 one recovers the Eulerian form of the usual guiding center equation
\[
q_h(\bE^*+{\mathbf{u}}_\textrm{gc}\times\bB^*)=m_h a_{\parallel\rm gc}\bb
\,,
\]
with the definitions \eqref{EffFields} and the relation \eqref{effE}.
In turn, this returns the 
phase-space vector field 
$\boldsymbol{\mathcal{X}}_\textrm{gc}=\big({\mathbf{u}}_\textrm{gc},a_{\parallel\text{gc}}\big)
$ as in \eqref{EulerianVF}, thereby producing the drift-kinetic equation \eqref{multi-fluid-Vlasov} for the phase-space density $F(\z,t;\mu)$, as in the second of \eqref{advection1} (upon replacing $\boldsymbol{\mathcal{X}}$ by $\boldsymbol{\mathcal{X}}_\textrm{gc}$). Then, equation \eqref{helpfulrel} still holds and the  fluid equation  \eqref{EP-CCS1} with \eqref{hotcurrgc} returns \eqref{cc-hybrid-momentum-GC}. To conclude the present formulation, \eqref{cc-hybrid-end-GC} arises by taking the curl of the third in \eqref{advection1}, while \eqref{cc-hybrid-mass-DK} and \eqref{J+M} follow upon defining $f=F/B^*_\|$ in the second of \eqref{advection1}. 

Although drift-kinetic equations are widely used in hybrid kinetic-MHD models, gyrokinetic theory is more suitable when the drift approximation does not hold. For example, the CCS in \cite{BeDeCh} makes use of gyrokinetic equations to formulate a gyrokinetic-MHD hybrid theory. The following sections are dedicated to this particular topic.

\section{Gyrokinetic-MHD\label{sec:GK-MHD}}
\subsection{Overview of hybrid gyrokinetic-MHD modeling\label{four_point_0ne}}
A more sophisticated low-frequency approximation of the Vlasov equation is provided by gyrocenter theory. Like guiding center theory, gyrocenter theory provides an efficient model of particle dynamics in a strongly magnetized plasma. The difference between gyrocenter theory and guiding center theory lies in the assumptions the two theories make regarding the dynamical electromagnetic field. Most generally, guiding center theory makes the following assumptions.
\begin{itemize}
\item[(GCA)] The parallel electric field scales as $\rho_c/L\ll1$, where $\rho_c$ is the gyroradius and $L=B/|\nabla B|$ is the magnetic scale length.

\item[(GCB)] Both the electric and magnetic field vary slowly in space and time relative to the gyroradius  and gyrofrequency, respectively.
\end{itemize}
In the version of guiding center theory used earlier in this work, it is also assumed that the perpendicular electric field scales as $\rho_c/L$. In contrast, gyrocenter theory makes the following alternative assumptions.
\begin{itemize}
\item [(GYA)] The electric and magnetic fields are given by $\bE(\bx,t)=\tilde{\bE}(\bx,t)$ and $\bB(\bx,t)=\bB_{\text{eq}}(\bx)+\tilde{\bB}(\bx,t)$, where $\tilde{\bE}$ and $\tilde{\bB}$ scale as $\rho_c/L_{\text{eq}}\ll1$ with $L_{\text{eq}}=B_{\text{eq}}/|\nabla B_{\text{eq}}|$.

\item [(GYB)] The parallel scale lengths and time scales of $\tilde{\bE}$ and $\tilde{\bB}$ are large compared to $\rho_c$ and the gyrofrequency, while the perpendicular scale lengths are comparable to $\rho_c$. 
\end{itemize}
The relaxed assumption on the perpendicular scale length of the fluctuating electromagnetic field is responsible for the additional theoretical sophistication present in gyrocenter theory. Any model that couples a distribution of gyrocenters to electromagnetic fields and perhaps additional plasma species is a gyrokinetic model. {\color{black}See
\cite{BrHa} for a detailed description of the difference between a particle's gyrocenter and guiding center.}

In this section, we will discuss hybrid models that couple a hot kinetic ensemble of gyrocenters to an MHD fluid, i.e. hybrid gyrokinetic-MHD (GK-MHD) models. Like the drift-kinetic-MHD (DK-MHD) models discussed earlier, these can be classified as being either current coupling or pressure coupling. Keeping with the general theme of this article, we will focus on the current coupling approach. As explained in \cite{PaBeFuTaStSu}, GK-MHD models can be regarded as ``particle closures" of the exact plasma momentum equation that enable the incorporation of important nonlinear wave-particle effects and kinetic turbulence effects into an otherwise fluid-based model.  

A terse account of the rationale behind the derivation of current-coupling GK-MHD models is given in \cite{ParkEtAl}. For the reader's convenience, we will now give a detailed derivation. We begin with \eqref{multi-fluid-momentum1}-\eqref{GaussGC1}, but with the Vlasov equation \eqref{multi-fluid-Vlasov1} replaced by its gyrocenter approximation,
\begin{gather}
\partial_t F+\nabla\cdot(F\mathbf{u}_{\text{gy}})+\partial_{v_\parallel}( F a_{\parallel \text{gy}})=0,\label{gyrocenter_continuity}
\end{gather}
where $F=F(\bX,v_\parallel)$ is the gyrocenter distribution function,  $\mathbf{u}_{\text{gy}}$ is the gyrocenter drift velocity, and $a_{\parallel\text{gy}}$ is the gyrocenter parallel acceleration.  As soon as a fluid equation of state is chosen {\color{black}(recall that in this work we adopt a barotropic equation of state)}, the gyrocenter equations of motion are specified, and the hot charge and current densities are related to the gyrocenter distribution function, this system is closed in the sense that there is the same number of equations as unknowns, {\color{black} i.e. $(\bU,\rho,\bB,F)$.}

As we have done in previous sections, to pass from this multi-fluid hybrid system to a current-coupling GK-MHD model, we apply the usual assumptions underlying MHD. The displacement current is dropped from Amp\`ere's law, and quasineutrality is used in place of Gauss's law. Then the sum of the fluid momentum equations is used to obtain a single-fluid momentum equation, while the dominant balance of the electron momentum equation is assumed to be given by ideal Ohm's law.
Taken together, these ``MHD assumptions" lead to the current-coupling GK-MHD model
\begin{align}
&\rho(\partial_t\bU+\bU\cdot\nabla\bU)=-q_h n_h\bE+(\bJ-\bJ_h)\times\bB-\nabla\mathsf{p}\label{hybrid_momentum_gen}\\
&\partial_t\rho+\nabla\cdot(\rho\bU)=0\\
&\partial_t F+\nabla\cdot(F\mathbf{u}_{\text{gy}})+\partial_{v_\parallel}(F a_{\parallel\text{gy}})=0\\
&\nabla\times\bB=\mu_0\bJ\\
&\partial_t\bB=-\nabla\times\bE\\
&\bE+\bU\times\bB=0\label{Ohmsl}\\
&\nabla\cdot\bB=0.\label{hybrid_divB_gen}
\end{align}
Note that Faraday's law and Ohm's law can be used in conjunction to eliminate the electric field $\bE$ from this system of equations. Like the multi-fluid-gyrokinetic hybrid model, as soon as a single-fluid equation of state is chosen , the gyrocenter equations of motion are given, and the hot charge and current densities are related to the gyrocenter distribution function, this hybrid gyrokinetic-MHD system has the same number of equations as unknowns.

\subsection{The current-coupling model of Belova, Denton, and Chan}

The most complete nonlinear current-coupling GK-MHD modeling effort is reported in \cite{BeDeCh}.  We will refer to the model in that paper as the BDC model for brevity. In the BDC model, the hot charge and current densities are expressed as
\begin{align}
q_hn_{h}(\bx)&= q_h\iint_\mu \langle\delta(\bX+\brho-\bx)\rangle F\,\de^4\z\\
\bJ_{h}(\bx)&= q_h\iint_\mu \langle(\mathbf{u}_{\text{gy}}+\omega_c\brho\times\bb_{\text{eq}})\delta(\bX+\brho-\bx)\rangle F\,\de^4\z,
\end{align}
where $\z=(\bX,v_\parallel)$ denotes the gyrocenter phase space coordinates, $\de^4\z=\de^3\bX\,dv_\parallel$, and 
\begin{align}
\brho&=q_h^{-1}\sqrt{2\mu m_h B_{\text{eq}}^{-1}(\bX)}\bb_{\text{eq}}(\bX)\times(\cos\theta\,{\bf e}_1(\bX)-\sin\theta\,{\bf e}_2(\bX))\label{rho_def}\\
\omega_c&=q_h B_{\text{eq}}(\bX)m_h^{-1}\\
\langle\,\cdot\,\rangle&=\frac{1}{2\pi}\int_0^{2\pi}\!\cdot\ d\theta,
\end{align}
with ${\bf e}_1,{\bf e}_2$ orthogonal fields of unit vectors spanning the plane perpendicular to $\bB_{\text{eq}}$. {\color{black} Here $\brho$ is the leading-order gyroradius vector, $\omega_c$ is the cyclotron frequency, and $\theta$ is the gyrophase.} Notice that one has  $\partial_\theta\brho=\brho\times\bb_{\text{eq}}$, where $\bb_{\text{eq}}=\bb_{\text{eq}}(\bX)$ unless otherwise specified. These expressions approximate the familiar particle-space current in terms of gyrocenter variables. For the gyrocenter dynamics, the BDC model uses equations inspired by, but not equivalent to equations in \cite{Brizard_JPP}, which derives gyrocenter equations of motion from a gyrocenter Lagrangian. Specifically, {\color{black}in the BDC model,}
\begin{align}
a_{\parallel\text{gy}}&=\frac{q_h}{m_h}\frac{\bB^{**}}{B_{\parallel}^{**}}\cdot\bE^{**}\label{BDC_apar}\\
\mathbf{u}_{\text{gy}}&=\frac{1}{B_{\parallel}^{**}}\bigg[\bB_{}^{**}v_\parallel+ \bE^{**}\times\bb_{\text{eq}}\bigg],\label{BDC_ugy}
\end{align}
where
\begin{equation}
\bB^{**}=\bB_{\text{eq}}+\langle\tilde{\bB}(\bX+\brho)\rangle
\,,\,\qquad\quad\ \\
\bE^{**}=\langle\tilde{\bE}(\bX+\brho)\rangle-q_h^{-1}\nabla([\mu+\delta\mu] B_{\text{eq}})
\end{equation}
and 
\[
B_{\parallel}^{**}=\bb_{\text{eq}}\cdot\bB_{}^{**}
\,.
\]
Here $\delta\mu$ is proportional to the fluctuating magnetic flux through the gyro orbit of a gyrocenter at $\bX$. Specifically,
\begin{align}
\delta\mu &=-q_h B_{\text{eq}}^{-1}\langle\omega_c\brho\times\bb_{\text{eq}}\cdot\tilde{\bA}(\bX+\brho)\rangle\nonumber\\
&=\frac{q_h^2}{2\pi m_h }\int_{D(\bX)} \tilde{\bB}\cdot d\mathbf{S},
\label{niceidentity}
\end{align}
where $D(\bX)$ is the disc swept out by $\bX+\brho$ as $\theta$ is increased from $0$ to $2\pi$, and $d\mathbf{S}=\bb_{\text{eq}}(\bX)\,dS$. Finally, the BDC model adopts an adiabatic equation of state, which we will replace in this discussion with the barotropic equation of state $\mathsf{p}=\mathsf{p}(\rho)$, where $\mathsf{p}(\rho)=\rho^2\,\mathcal{U}'(\rho)$, and $\mathcal{U}$ is the single-fluid internal energy density.

Note that it appears the authors of \cite{BeDeCh} were unaware of the second equality in \eqref{niceidentity},
which follows from Stoke's theorem. This identity proves that the gyrocenter equations of motion in the BDC model can always be written without the fluctuating vector potential $\tilde{\bA}$ appearing explicitly, even when the background magnetic field is non-uniform. While this does contradict a claim in \cite{BeDeCh}, it also puts the BDC model on a stronger theoretical footing.

While formulating this model, the authors of \cite{BeDeCh} paid special attention to the model's conservative properties. They claimed that the BDC model has exact energy and momentum conservation laws. However, the discussion in \cite{BeDeCh} does not indicate if these conservation laws are only valid when the background is uniform, or if they are valid for more general sorts of background fields. We have checked a number of conservation laws that one would want the BDC model to satisfy in order to clarify this issue. 
\begin{itemize}
\item {\bf Energy is conserved:} Assuming periodic boundary conditions, the time derivative of the system energy,
\begin{align}\label{GK-energy}
E=\iint_\mu\bigg(\frac{1}{2}m_hv_\parallel^2+[\mu+\delta\mu]B_{\text{eq}}\bigg)F\,\de^4\z\,+\int\bigg(\frac{1}{2}\rho|\bU|^2-\rho\,\mathcal{U}(\rho)-\frac{1}{2\mu_0}|\bB|^2\bigg)\,\de^3\bx,
\end{align}
is in fact zero regardless of the form of the background magnetic field. 

\item {\bf Momentum is not conserved:} The total momentum
\begin{align}
\mathbf{N}=\iint_\mu m_hv_\parallel\bb_{\text{eq}}\,F\,\de^4\z+\int\!\rho\,\bU\,\de^3\bx,\label{belova_mom}
\end{align}
satisfies
\begin{multline}
\frac{d\mathbf{N}}{dt}=\iint_\mu\bigg(m_h v_\parallel\mathbf{u}_{\text{gy}}\cdot\nabla\bb_{\text{eq}}-q_h \mathbf{u}_{\text{gy}}\times\langle\Delta\bB\rangle-q_h \langle\omega_c(\brho\times\bb_{\text{eq}})\times\Delta\bB \rangle\nonumber\\
-\nabla([\mu+\delta\mu]B_{\text{eq}})-q_h\langle\omega_c(\brho\times\bb_{\text{eq}})\times\tilde{\bB}(\bX+\brho)\rangle\bigg)F\,\de^4\z,
\end{multline}
where $\Delta\bB=\bB_{\text{eq}}(\bX+\brho)-\bB_{\text{eq}}(\bX)$, which apparently only vanishes when the background magnetic field is {\color{black}uniform (i.e. constant in space).} 

\item {\bf Phase space volume is not conserved:} The rate at which phase space volume increases is given by
\begin{multline}\label{GK-Liouville}
\partial_t B_{\parallel}^{**}+\nabla\cdot(B_{\parallel}^{**}\mathbf{u}_{\text{gy}})+\partial_{v_\parallel}(B_{\parallel}^{**}a_{\parallel\text{gy}})=\bb_{\text{eq}}\cdot[\nabla\times\langle\tilde{\bE}(\bX+\brho)\rangle-\langle(\nabla\times\tilde{\bE})(\bX+\brho)\rangle]\\
+v_\parallel\nabla\cdot\langle\tilde{\bB}(\bX+\brho)\rangle-\bE^{**}\cdot\nabla\times\bb_{\text{eq}},
\end{multline}
which also only vanishes when the background field is {\color{black}uniform}.

\item {\bf Hot charge is not conserved:} The hot charge density satisfies 
\begin{align}\label{GK-continuity}
\partial_tq_h n_h+\nabla\cdot\bJ_h&=\iint_\mu q_h\langle(\mathbf{u}_{\text{gy}}+\omega_c\brho\times\bb_{\text{eq}})\cdot\nabla\brho\cdot(\nabla\delta)(\bX+\brho-\bx)\rangle F\,\de^4\z,
\end{align}
which shows that the hot charge in the BDS model is only conserved when the background magnetic field is {\color{black}uniform}. Note that the right-hand-side of this equation is not a divergence, which implies that charge conservation breaks both locally and globally.
\end{itemize}

Altogether, the above expressions demonstrate that the BDC model enjoys a number of exact conservation laws when the background magnetic field is {\color{black}uniform}, but, as soon as background inhomogeneity is introduced, each of these laws except the one for energy is broken.

\subsection{Formulation of the new gyrokinetic-MHD model}
The conservation laws that are broken by the BDC model in an inhomogeneous background are obvious physical constraints that any sound hybrid model should satisfy. Moreover, background magnetic nonuniformity is crucial to include in any realistic model of hot particle effects in magnetic fusion devices; the inhomogeneity is responsible for a number of important wave-particle resonance effects. We are therefore compelled to formulate a new GK-MHD model that possesses the good conservation properties of the {\color{black}uniform}-background BDC model, regardless of the form of the background magnetic field. In this subsection, we will formulate such a model while modifying the BDC model as little as possible. 

As discussed in Section \ref{four_point_0ne}, we may specify a GK-MHD model by giving expressions for the hot charge and current densities, as well as the gyrocenter equations of motion. In our new GK-MHD model, we will express the charge and current densities as
\begin{align}
q_hn_h(\bx)=&\ q_h\iint_\mu  \langle\delta(\bX+\brho-\bx)\rangle F\,\de^4\z\label{new_hot_charge}\\
\bJ_h(\bx)=&\iint_\mu  q_h\langle(\mathbf{u}_{\text{gy}}+\omega_c\brho\times\bb_{\text{eq}})\delta(\bX+\brho-\bx)\rangle F\,\de^4\z\nonumber\\
&+\iint_\mu  q_h\langle\!\langle\mathbf{u}_{\text{gy}}\cdot(\nabla_{\bX}+\nabla_{\bx})[\delta(\bX+\lambda\brho-\bx)\brho]\rangle\!\rangle F\,\de^4\z,\label{new_hot_current}
\end{align}
and the gyrocenter equations of motion as
\begin{align}
a_{\parallel\text{gy}}&=\frac{q_h}{m_h}\frac{\bB^*}{B_\parallel^*}\cdot\bE^*\label{new_apar}\\
\mathbf{u}_{\text{gy}}&=\frac{v_\parallel \bB^*}{B_\parallel^*}+\frac{\bE^*\times\bb_{\text{eq}}}{B_{\parallel}^*},\label{new_ugy}
\end{align}
where
\begin{align}
\bB^*=&\ \bB(\bX)+m_hq_h^{-1}v_\parallel\nabla\times\bb_{\text{eq}}+\nabla\times\langle\!\langle\tilde{\bB}(\bX+\lambda\brho)\times\brho\rangle\!\rangle\\
\bE^*=&\ \tilde{\bE}(\bX)-q_h^{-1}\nabla([\mu+\delta\mu]B_{\text{eq}})+\langle\!\langle(\nabla\times\tilde{\bE})(\bX+\lambda\brho)\times\brho\rangle\!\rangle\nonumber\\
&+\nabla\langle\!\langle\tilde{\bE}(\bX+\lambda\brho)\cdot\brho\rangle\!\rangle.
\end{align}
Note that by virtue of the $\nabla\times\bb_{\text{eq}}$ term in $\bB^*$, the first term in the right-hand-side of \eqref{new_ugy} captures the curvature drift, in contrast to the BDC model. The double angle brackets appearing in these expressions are defined as
\begin{align}\label{dab}
\langle\!\langle Q\rangle\!\rangle=\frac{1}{2\pi}\int_0^1\!\int_0^{2\pi}\!Q\,d\theta\,d\lambda ={\color{black} \int_0^1\langle Q\rangle\,d\lambda},
\end{align}
which were previously introduced \cite{Porazik_2011,BuBrMoQi} and implemented numerically in \cite{Porazik_2011} (\cite{BuBrMoQi} did not use the double bracket notation, but the same parameter $\lambda$ was introduced). {\color{black} Note that the variable $\lambda$ may be interpreted as a normalized radial coordinate on the disc swept out by rotating a particle's gyroradius vector.} The equations defining our model are therefore (\ref{hybrid_momentum_gen}-\ref{hybrid_divB_gen}) with $n_h$, $\bJ_h$, $a_{\parallel\text{gy}}$, and $\mathbf{u}_{\text{gy}}$ given by (\ref{new_hot_charge}-\ref{new_ugy}). 

It is not difficult to verify that this model agrees with the BDC model when the background magnetic field is {\color{black}uniform}. However, {\color{black}as a result of the modified gyrocenter dynamics and hot current density}, this model possesses the following conservation laws regardless of the form of the background magnetic field.  
\begin{itemize}

\item {\bf Energy is conserved:} Assuming for concreteness a barotropic equation of state for the MHD fluid, the system energy \eqref{GK-energy}
is conserved. Note that this energy is the same as the energy in the BDC model.

\item {\bf Momentum is conserved:} When the background magnetic field is either axisymmetric or translation symmetric, there is a corresponding total momentum that is conserved exactly. For instance, assuming the background field is symmetric under rotations about the $z$-axis, the total toroidal momentum
\begin{align}
N_\phi=&\iint_\mu m_h v_\parallel \bb_{\text{eq}}\cdot {\bf e}_z\times\bX\,F\,\de^4\z+\int \rho\,\bU\cdot {\bf e}_z\times\bx\,\de^3\bx\nonumber\\
&+\iint_\mu  q_h\langle\!\langle[\brho\times\bB_{\text{eq}}(\bX+\lambda\brho)]\cdot[{\bf e}_z\times\bX]\rangle\!\rangle F\,\de^4\z\nonumber\\
&+\iint_\mu q_h \langle\!\langle\lambda {\bf e}_z\cdot\brho\brho\cdot \bB_{{}}(\bX+\lambda\brho)-\lambda |\brho|^2 {\bf e}_z\cdot\bB_{{}}(\bX+\lambda\brho)\rangle\!\rangle F\,\de^4\z
\end{align}
is conserved exactly. If the background is invariant under arbitrary spatial translations (which implies the $\bB_{\text{eq}}$ is {\color{black}uniform}), then the momentum in Eq.\,(\ref{belova_mom}) is conserved. If the background is axisymmetric but not translation invariant, the total momentum is not conserved; the mechanical structures (e.g. coils) that produce the background field exert a force on the plasma while producing zero vertical torque.

\item {\bf Phase space volume is conserved:} Phase space volume is preserved exactly, so that \eqref{GK-Liouville} becomes
\begin{gather}
\partial_t B_{\parallel}^*+\nabla\cdot(B_{\parallel}^*\mathbf{u}_{\text{gy}})+\partial_{v_\parallel}(B_{\parallel}^*a_{\parallel\text{gy}})=0,
\end{gather}
which represents the time-dependent form of Liouville's theorem, and is the gyrocenter analogue of Eq.\,(\ref{liouville_gc}).

\item {\bf Hot charge is conserved:} Hot charge is conserved locally and \eqref{GK-continuity} becomes
\begin{gather}
q_h\partial_tn_h+\nabla\cdot\bJ_h=0.
\end{gather}

\end{itemize}
In addition, we remark that this new model preserves the standard expressions for  cross helicity, magnetic helicity, and the family of invariants $\int \!\Phi(F/B_\parallel^*)\,B_\parallel^*\,dv_\parallel\,d^3\bX$ (comprising entropy), as may be verified by a direct calculation in the same spirit of \cite{HoTr2011}.

This new GK-MHD model can be regarded as a corrected version of the BDC model. Both the hot current density and the gyrocenter equations of motion contain correction terms that are proportional to gradients of the background magnetic field. These correction terms ensure that the model obeys the conservation laws that are broken by the original BDC model. At the same time, they incorporate the effects of the curvature drift. However, at this stage it is entirely unclear how we chose our correction terms, or if they have any fundamental significance.
Therefore, in the next two sections we will carry out the task of deriving our new GK-MHD model from first principles. First we will digress briefly on a technical, but important aspect of gyrocenter motion in electromagnetic fields. Then we will use the hybrid variational method employed twice in this article already to systematically derive our new GK-MHD model.

\subsection{Gauge invariance in single-particle gyrocenter theory}
Because we aim to derive our new GK-MHD model from a variational principle, we must reckon with a basic fact about variational principles involving electromagnetic fields, and another basic fact about existing single-gyrocenter Lagrangians in the literature. The first fact {\color{black} \cite{Bleeker}}: variational principles that are symmetric under gauge transformations are consistent with particle-space charge conservation; variational principles without this symmetry are inconsistent with charge conservation. The second fact: {\color{black}many} truncated electromagnetic single-gyrocenter Lagrangians in the literature are not gauge invariant in the presence of background magnetic nonuniformity, even though the formal all-orders gyrocenter Lagrangian must be so. {\color{black} (A thorough investigation of which truncated single-gyrocenter Lagrangians in the literature are gauge invariant is the subject of ongoing work)}. Taken together, these two facts imply that we must identify a new truncated gauge invariant single-gyrocenter Lagrangian before we can derive a particle-space charge conserving variational GK-MHD model. The purpose of this section is to derive such a single-gyrocenter Lagrangian. For more details on variational principles for individual phase-space trajectories, see the Appendix.

As explained for instance in \cite{Brizard_JPP}, the equations of motion for gyrocenters in time-dependent electromagnetic fields should be given as the Euler-Lagrange equations associated with a single-gyrocenter Lagrangian. The single-gyrocenter Lagrangian that comes the closest to giving the gyrocenter equations of motion in the BDC model is 
\begin{multline}
\ell_{0}=\big(q_h\bA_{\text{eq}}+m_hv_\parallel \bb_{\text{eq}}\big)\cdot \dot{\bX}+q_h\langle\tilde{\bA}(\bX+\brho)\rangle\cdot\dot{\bX}\nonumber\\
-\bigg(\frac{1}{2}m_h v_\parallel^2+\mu B_{\text{eq}}+q_h\langle\tilde{\varphi}(\bX+\brho)\rangle-q_h\langle\omega_c\brho\times\bb_{\text{eq}}\cdot\tilde{\bA}(\bX+\brho)\rangle\bigg),
\end{multline}
which differs from the Lagrangian given in Eq.\,(10) of \cite{Brizard_JPP} in only one respect. 
While the $\brho$ appearing in this expression generally depends on $\bX$ through the background magnetic field (according to Eq.\,(\ref{rho_def})), the $\brho$ in \cite{Brizard_JPP} is assumed to be $\bX$-independent.  Note that when the background magnetic field is uniform so that $\bb_{\text{eq}}$ and $\brho$ are independent of $\bX$, this Lagrangian's Euler-Lagrange equations reproduce the expressions for $a_{\parallel\text{gy}}$ and $\mathbf{u}_{\text{gy}}$ in the BDC model, namely Eq.\,(\ref{BDC_apar}) and Eq.\,(\ref{BDC_ugy}). However, when the background magnetic field is not {\color{black}uniform}, these Euler-Lagrange equations do not match the BDC model's gyrocenter dynamical equations. 

Consider now the behavior of $\ell_{0}$ under gauge transformations. If $\tilde{\bA}$ is replaced with $\tilde{\bA}^\prime=\tilde{\bA}+\nabla \psi$, where $\psi$ is a time-independent scalar, then
\begin{align}
\ell_{0}^\prime=\ell_{0}+\frac{q_h}{c}\langle(\nabla \psi)(\bX+\brho)\rangle\cdot\dot{\bX}.
\end{align}
When the gradient of the background magnetic field is zero, $(\nabla \psi)(\bX+\brho)=\nabla(\psi(\bX+\brho))$, and it is readily seen that $\ell_{0}^\prime$ differs from $\ell_{0}$ by a total time derivative. This means $\ell_{0}$ is gauge invariant when the background magnetic field is {\color{black}uniform}. However, when the background field gradient is non-zero, $(\nabla \psi)(\bX+\brho)\neq\nabla(\psi(\bX+\brho))=([1+\nabla\brho]\cdot\nabla \psi)(\bX+\brho)$, and the difference between $\ell_{0}^\prime$ and $\ell_{0}$ is not a total time derivative. This indicates that the Euler-Lagrange equations associated with $\ell_{0}$ are gauge invariant \emph{only when the background field is uniform}. For similar reasons, {\color{black}many} other single-gyrocenter Lagrangians in the literature suffer from this same flaw.

In order to remedy this technical issue, we will modify the Lagrangian $\ell_{0}$ as follows. First we use the simple identity $Q(\bX+\brho)=Q(\bX)+\int_0^1\brho\cdot\nabla Q(\bX+\lambda\brho)\,d\lambda$ to rewrite $\ell_{0}$ as
\begin{multline}
\ell_{0}=\Big(q_h\bA_{\text{eq}}+m_hv_\parallel \bb_{\text{eq}}\Big)\cdot \dot{\bX}+q_h\tilde{\bA}(\bX)\cdot\dot{\bX}
-\bigg(\frac{1}{2}m_h v_\parallel^2+[\mu+\delta\mu] B_{\text{eq}}+q_h\tilde{\varphi}(\bX)\bigg)
\\
+q_h\langle\!\langle\brho\cdot(\nabla\tilde{\bA})(\bX+\lambda\brho)\cdot\dot{\bX}\rangle\!\rangle-q_h\langle\!\langle\brho\cdot(\nabla\tilde{\varphi})(\bX+\lambda\brho)\rangle\!\rangle.
\end{multline}
Next we subtract from $\ell_{0}$ a total time derivative,
\begin{align}
\ell_{0}\rightarrow \ell_{0}-\frac{d}{dt}q_h\langle\!\langle\tilde{\bA}(\bX+\lambda\brho)\cdot\brho\rangle\!\rangle,
\end{align} 
{\color{black}where we recall the definition \eqref{dab}.} Note that at this stage the single-gyrocenter Lagrangian is still equivalent to $\ell_{0}$. Finally, we replace the total time derivative with its approximation
\begin{align}
\frac{d}{dt}q_h\langle\!\langle\tilde{\bA}(\bX+\lambda\brho)\cdot\brho\rangle\!\rangle\approx q_h\langle\!\langle\dot{\bX}\cdot\nabla\tilde{\bA}(\bX+\lambda\brho)\cdot\brho+\partial_t\tilde{\bA}(\bX+\lambda\brho)\cdot\brho\rangle\!\rangle,
\end{align}
where the neglected terms are proportional to gradients of the background magnetic field. The resulting {\color{black}single} gyrocenter Lagrangian {\color{black}now replacing $
ell_0$} can be expressed as
\begin{multline}
\ell_{\text{gy}}\equiv\Big(q_h\bA_{\text{eq}}+m_h v_\parallel\bb_{\text{eq}}\Big)\cdot\dot{\bX}
-\bigg(\frac{1}{2}m_hv_\parallel^2+[\mu+\delta\mu] B_{\text{eq}}\bigg)
\\
+q_h\tilde{\bA}(\bX)\cdot\dot{\bX}-q_h\tilde{\varphi}(\bX)
+q_h\langle\!\langle[\tilde{\bE}(\bX+\lambda\brho)+\dot{\bX}\times\tilde{\bB}(\bX+\lambda\brho)]\cdot\brho\rangle\!\rangle.\label{new_gy_lag}
\end{multline}
In terms of the gyrocenter asymptotic expansion (see \cite{Brizard_JPP}), $\ell_{\text{gy}}$ is just as accurate as $\ell_{0}$. However, $\ell_{\text{gy}}$ only changes by the addition of a total time derivative when the potentials are subject to a gauge transformation. Thus, the Euler-Lagrange equations associated with $\ell_{\text{gy}}$ are gauge invariant, in contrast to those associated with $\ell_{0}$. In Section \ref{sec:var_GK}, we will use $\ell_{\text{gy}}$ to construct the particle contribution to a GK-MHD system Lagrangian. 
\rem{ 
An alternative approach to restoring gauge invariance in $\ell_0$ is to recognize that the term $\langle\omega_c\brho\times\bb\cdot \tilde{\bA}(\bX+\brho)\rangle$ in $\ell_0$ is merely the lowest-order approximation to $\langle\bv_\perp\cdot\tilde{\bA}(\bX+\brho)\rangle$, where $\bv_\perp$ is the particle-space perpendicular velocity. In guiding center theory, while the lowest-order perpendicular velocity is $\omega_c \brho\times\bb_{\text{eq}}$, the first-order perpendicular velocity is
\begin{align}
\bv_\perp\approx \omega_c\brho\times\bb_{\text{eq}}+\dot{\bX}\cdot\nabla\brho,\label{vperp_mod}	
\end{align}
\comment{CT: this notation is inconsistent}\noindent
which is consistent with Eq.\,(3.17) of \cite{CaBr}.
It therefore makes physical sense to make the replacement
\begin{align}
\langle\omega_c\brho\times\bb_{\text{eq}}\cdot\tilde{\bA}(\bX+\brho)\rangle\rightarrow \langle[\omega_c\brho\times\bb_{\text{eq}}+\dot{\bX}\cdot\nabla\brho]\cdot\tilde{\bA}(\bX+\brho)\rangle.	
\end{align}
If this modification is made in $\ell_0$, the resulting Lagrangian is equal to $\ell_{\text{gy}}$ modulo gauge-invariant terms that are proportional to products of the fluctuating magnetic field with gradients of the background magnetic field.
} 

The procedure used here to ``repair" an existing gyrocenter Lagrangian could, in principle, be applied to any gyrocenter Lagrangian in the literature.  However, repairing a higher-order gyrocenter Lagrangian is a much more cumbersome task than the procedure used in this section might suggest. A more powerful and systematic method for deriving an all-orders gauge invariant gyrocenter Lagrangian will be presented in a future publication, {
\color{black} along with the method's physical underpinnings.} The first results obtained using this systematic theory are reported in \cite{BuBrMoQi}.

\subsection{The variational framework for gyrokinetic-MHD\label{sec:var_GK}}
We will now present a first-principles variational derivation of our new GK-MHD model. In order to construct a Lagrangian for a hybrid system consisting of hot particles and an MHD fluid, we should sum the Lagrangians corresponding to the two subsystems. The first of these subsystems is the gyrocenter ensemble, whose Lagrangian is given by
\begin{align}
L_{\text{p}}=&
\iint_\mu \left[\Big(q_h\bA_{\text{eq}}+m_h v_\parallel\bb_{\text{eq}}\Big)\cdot\mathbf{u}_{\text{gy}}(\bz)- \frac{1}{2}m_hv_\parallel^2+[\mu+\delta\mu] B_{\text{eq}}\right]\,F\,\de^4\z\nonumber\\
&
+\iint_\mu q_h\Big(\tilde{\bA}(\bX)\cdot\mathbf{u}_{\text{gy}}(\bz)-\langle\!\langle\mathbf{u}_{\text{gy}}(\bz)\cdot\brho\times\tilde{\bB}(\bX+\lambda\brho)\rangle\!\rangle\Big)F\,\de^4\z\nonumber\\
&
+\iint_\mu q_h\Big(\langle\!\langle\brho\cdot\tilde{\bE}(\bX+\lambda\brho)\rangle\!\rangle-\tilde{\varphi}(\bX)\Big)F\,\de^4\z,
\label{general_particle_lagrangian}
\end{align}
where the integrand is the Eulerian (Euler-Poincar\'e) version of $\ell_{\text{gy}}$ (see Eq.\,(\ref{new_gy_lag})), which can be written upon replacing $\dot{\bX}$ with $\mathbf{u}_{\text{gy}}(\bz)$ (cf. \cite{SqQiTa}). 
Here $\bcalX_{\text{gy}}=(\mathbf{u}_{\text{gy}},a_{\parallel\text{gy}})$ is the Eulerian phase space fluid velocity. Note that this expression is consistent with the Appendix, which describes generally how a single-particle Lagrangian (in this case $\ell_{\text{gy}}$) can be used to construct a Lagrangian for a collisionless distribution of particles. The second subsystem is the barotropic MHD fluid, whose Lagrangian is given by \eqref{MHD-Lagrangian},
where we recall that $\bA=\bA_{\text{eq}}+\tilde{\bA}$ and we note that there would be no essential difficulty were $\mathcal{U}$ to depend on both density and entropy. The hybrid system Lagrangian is therefore $L=L_{\text{p}}+L_{\text{MHD}}$, and the hybrid variational principle is written as in \eqref{VarPrinc}, 
where $S_{}=\int L_{}\,dt$ is the hybrid action and the precise meaning of the symbol $\delta$ will now be described.

The hybrid action $S_{}$ may be regarded as a functional of paths through $(\boldsymbol{z},\boldsymbol{q})$-space, where $\boldsymbol{z}(\bz_0)=(\boldsymbol{X}(\bz_0),\mathcal{V}_\parallel(\bz_0))$ is the gyrocenter phase space fluid configuration map and $\boldsymbol{q}(\bx_0)$ is the MHD fluid configuration map. 
In order to see this, first observe that hot particle and mass conservation imply the gyrocenter distribution function and the MHD mass density are related to their initial values according to the first two in \eqref{pullbacks}, that is
\begin{align}
F(\boldsymbol{z}(\bz_0))\,\de^4\boldsymbol{z} &=F_0(\bz_0)\,\de^4\z_0
\,,\qquad\quad\ 
\rho(\boldsymbol{q}(\bx_0))\,\de^3\boldsymbol{q} =\rho_0(\bx_0)\,\de^3\bx_0.
\end{align}
Next, note that the Eulerian fluid velocities $(\bcalX_{\text{gy}},\bU)$ are given in terms of the fluid configuration maps $(\boldsymbol{z},\boldsymbol{q})$ according to
\begin{align}
\bcalX_{\text{gy}}(\boldsymbol{z}(\bz_0))&=\dot{\boldsymbol{z}}(\bz_0)\label{gk_phase_fluid}
\end{align}
	and Eq.\,(\ref{configuration_space_velocity_def}), i.e. $\mathbf{u}_{\text{gy}}(\boldsymbol{z}(\bz_0))=\dot{\boldsymbol{X}}(\bz_0)$, $a_{\parallel\text{gy}}(\boldsymbol{z}(\bz_0))=\dot{\mathcal{V}}_\parallel(\bz_0)$, {\color{black}where we have suppressed the time dependence for notational simplicity}. Note that Eq.\,(\ref{gk_phase_fluid}) is merely the gyrocenter version of Eq.\,(\ref{phase_space_velocity_def}).
The relations just presented suffice to express $F,\rho,\bcalX_{\text{gy}},\bU$ in terms of the maps $\boldsymbol{z}$ and $\boldsymbol{q}$. In order to express $\bA$ and $\varphi$ in terms of the configuration maps, we must invoke the ideal Ohm's law\,(\ref{Ohmsl}). To ensure that the total magnetic field is advected by the bulk flow (a property implied by the curl of Ohm's Law), we will freeze the total vector potential as follows:
\begin{align}
(\tilde{\bA}(\boldsymbol{q}(\bx_0))+\bA_{\text{eq}}(\boldsymbol{q}(\bx_0)))\cdot d\boldsymbol{q} &=(\tilde{\bA}_0(\bx_0)+\bA_{\text{eq}}(\bx_0))\cdot d\bx_0.
\label{total_advection}
\end{align}
Then, in order to guarantee that Ohm's Law is completely satisfied, we will suppose, as before, that $\varphi$ is expressed in the hydrodynamic gauge \eqref{hygauge} as follows:
\begin{align}
\tilde{\varphi}&=
\bA\cdot\bU
=
(\bA_{\text{eq}}+\tilde{\bA})\cdot\bU.
\label{hydro_gauge}
\end{align}
Note that 
Eq.\,(\ref{total_advection}) implies the total vector potential is advected by the bulk while its fluctuating part is not. Also note that\,(\ref{total_advection})-(\ref{hydro_gauge}) give $\tilde{\bA}$ and $\tilde{\varphi}$ in terms of the fluid configuration map $\boldsymbol{q}$. Finally, the formulas $\tilde{\bE}=-\partial_t\tilde{\bA}-\nabla\tilde{\varphi}$ and $\tilde{\bB}=\nabla\times\tilde{\bA}$ imply expressions for the electric and magnetic field in terms of $\boldsymbol{q}$. Thus, we have shown that all terms in the hybrid Lagrangian can be expressed in terms of the fluid configuration maps $\boldsymbol{z},\boldsymbol{q}$ and their (first) time derivatives.

We may now sensibly vary the hybrid action $S$ by 
applying Euler-Poincar\'e reduction theory\,\cite{HoMaRa1998} and by recalling the relationships given in the previous paragraph: this amounts to varying $S$ using the following constrained variations
\begin{align}
&\delta F=-\nabla\cdot(F\bXi_{\bX})-\partial_{v_\parallel}(F\Xi_{v_\parallel})=-\nabla_{\z}\cdot(F\bXi)\\
&\delta\rho=-\nabla\cdot(\rho\bxi)\\
&\delta\bcalX_{\text{gy}}=\partial_t{\bXi}+\bcalX_{\text{gy}}\cdot\nabla_{\z}\bXi-\bXi\cdot\nabla_{\z}\bcalX_{\text{gy}}\\
&\delta\bU=\partial_t{\bxi}+\bU\cdot\nabla\bxi-\bxi\cdot\nabla\bU\\
&\delta\tilde{\bA}=\bxi\times\bB-\nabla(\bxi\cdot\bA)\label{tildeA_variation}
\end{align}
where $\bXi(\boldsymbol{z}(\bz_0))=\delta\boldsymbol{z}(\bz_0)=(\bXi_{\bX}(\boldsymbol{z}(z_0)),\Xi_{v_\parallel}(\boldsymbol{z}(\bz_0)))$ is the Eulerian phase space fluid displacement vector, and $\bxi(\boldsymbol{q}(\bx_0))=\delta\boldsymbol{q}(\bx_0)$ is the Eulerian MHD fluid displacement vector. Note that the first four of these relations are equivalent to the corresponding relations in Eqs.\,(\ref{variations-A}-\ref{variations1}), while Eq.\,(\ref{tildeA_variation}) differs from the corresponding relation in (\ref{variations1}) due to the presence of a background magnetic field. For the reader's convenience, we remark that these constrained variations also imply the relations
\begin{align}
\delta\tilde{\varphi} &=-\bxi\cdot\nabla\tilde{\varphi}+\partial_t{\bxi}\cdot\bA\\
\delta\tilde{\bB} &=\nabla\times(\bxi\times\bB)\\
\delta\tilde{\bE} &=\bxi\times(\nabla\times\tilde{\bE})-\nabla(\bxi\cdot\tilde{\bE})-(\partial_t{\bxi})\times\bB.	
\end{align}
 Using these relationships, the first variation of the hybrid Lagrangian can be computed as
\begin{align}
\delta L=&
\iint_\mu \left[\bXi_{\bX}\cdot\bigg(q_h\bE^*-q_h\bB^*\times\mathbf{u}_{\text{gy}}-m_h a_{\parallel\text{gy}}\bb_{\text{eq}}\bigg)+\Xi_{v_\parallel}\bigg(m_h\mathbf{u}_{\text{gy}}\cdot\bb_{\text{eq}}-m_h v_\parallel\bigg)\right]F\,\de^4\z\nonumber\\
&-\int\bigg(\rho(\partial_t\bU+\bU\cdot\nabla\bU)+\nabla \mathsf{p}- (\bJ-\bJ_h)\times\bB+q_hn_h\bE\bigg)\cdot\bxi\,\de^3\bx
\label{GY-vars}
\end{align}
where we recall the {\color{black}expression} $\mathsf{p}=\rho^2\,\mathcal{U}'(\rho)$ for the MHD fluid pressure as well as the usual relation $\mu_0\bJ=\nabla\times\bB$. {\color{black}In \eqref{GY-vars}, we have omitted the boundary terms
\begin{align}
&\frac{\de}{\de t}\int(\rho\bU-q_hn_{\text{gy}}\bA+\bP_h\times\bB)\cdot\bxi\,\de^3\bx\nonumber\\
&+\frac{\de}{\de t}\iint_\mu\bXi_{\bX}\cdot\bigg(q_h\bA+m_hv_\parallel\bb_{\text{eq}}+q_h \langle\!\langle\tilde{\bB}(\bX+\lambda\brho)\times\brho\rangle\!\rangle\bigg) F\,\de^4\z
\end{align}
emerging from integration by parts. These terms can be used to quickly verify conservation laws via Noether's theorem}. In addition, the hot charge and current densities are given by
\begin{align}
q_h n_h&=q_h\iint_\mu F(\bx,v_\parallel)\,dv_\parallel-\nabla\cdot\bP_{\text{gy}}\nonumber\\
&\equiv q_h n_{\text{gy}}-\nabla\cdot\bP_{\text{gy}}\label{total_charge_dielectric}\\
\bJ_h&=q_h\iint_\mu \mathbf{u}_{\text{gy}}(\bx,v_\parallel)\,F(\bx,v_\parallel)\,dv_\parallel+\nabla\times\bM_{\text{gy}}+\partial_t\bP_{\text{gy}}\nonumber\\
&\equiv \bJ_{\text{gy}}+\nabla\times\bM_{\text{gy}}+\partial_t\bP_{\text{gy}},\label{total_current_dielectric}
\end{align}
with the gyrocenter ensemble's polarization and magnetization densities given by (see \cite{YeKaufman} and more recently in \cite{BuBrMoQi})
\begin{align}\label{varders}
\bP_{\text{gy}} &=\frac{\delta L_{\text{p}}}{\delta\widetilde{\bE}}
\,,\qquad\quad\,
\bM_{\text{gy}} =\frac{\delta L_{\text{p}}}{\delta\widetilde{\bB}}.	
\end{align}
Indeed, note that Eqs.\,(\ref{total_charge_dielectric})-(\ref{total_current_dielectric}) are consistent with the well-known fact\,\cite{Krommes} that an ensemble of gyrocenters behaves as a polarized magnetized medium. 

Because $\bXi$ and $\bxi$ are arbitrary and we may assume $F$ and $\rho$ are nowhere vanishing, the variational principle $\delta S_{}=0$ is satisfied if and only if 
\begin{align}
&q_h\bE^*-q_h\bB^*\times\mathbf{u}_{\text{gy}}-m_h a_{\parallel\text{gy}}\bb_{\text{eq}}=0\label{gk_ham}\\
&m_h\mathbf{u}_{\text{gy}}\cdot\bb_{\text{eq}}-m_h v_\parallel=0\label{gk_ham2}\\
&\rho(\partial_t\bU+\bU\cdot\nabla\bU)=-\nabla \mathsf{p}-q_h n_h\bE+(\bJ-\bJ_h)\times\bB.\label{gk_mom}
\end{align}
These equations must be supplemented by the evolution equations that are implicitly built into the hybrid variational principle, namely
\begin{align}
&\partial_t F+\nabla\cdot(F\mathbf{u_{\text{gy}}})+\partial_{v_\parallel}(F a_{\parallel\text{gy}})=0\label{gk_vlasov}\\
&\partial_t{\rho}+\nabla\cdot(\rho\,\bU)=0\label{gk_cont}\\
&\bE+\bU\times\bB=0\label{gk_0hm}.
\end{align}

Equations\,(\ref{gk_ham})-(\ref{gk_mom}), and (\ref{gk_vlasov})-(\ref{gk_0hm}) recover our new GK-MHD model. To see this, first observe that the functional derivatives \eqref{varders} defining the polarization and magnetization densities can be computed explicitly, giving
\begin{align}
\bP_\text{gy}(\bx)&= q_h\iint_\mu \langle\!\langle\delta(\bX+\lambda\brho-\bx)\brho\rangle\!\rangle F\,\de^4\z\\
\bM_\text{gy}(\bx)&= q_h \iint_\mu\langle\!\langle\delta(\bX+\lambda\brho-\bx)\brho\times[\mathbf{u}_{\text{gy}}+\lambda\omega_c\partial_\theta\brho]\rangle\!\rangle F\,\de^4\z.
\end{align}
Substituting these expressions into Eqs.\,(\ref{total_charge_dielectric})-(\ref{total_current_dielectric}), it is straightforward to verify that Eqs.\,(\ref{new_hot_charge})-(\ref{new_hot_current}) are recovered. Finally, it is also straightforward to verify that Eqs.\,(\ref{gk_ham})-(\ref{gk_ham2}) reproduce Eqs.\,(\ref{new_apar})-(\ref{new_ugy}).

This derivation of our new GK-MHD model makes one of the model's basic conservation laws very clear. Local conservation of hot charge is an immediate consequence of Eq.\,(\ref{total_charge_dielectric}-\ref{total_current_dielectric}). On the other hand, the validity of the global conservation laws for energy and momentum is not as clear. In fact all of these conservation laws can be seen as consequences of Noether's theorem. Energy and momentum conservation follow from time translation invariance and rotation invariance of the hybrid Lagrangian. Likewise, hot charge conservation follows from gauge invariance of the hybrid Lagrangian. 

\section{Conclusions}
In this paper we have introduced two new nonlinear variational hybrid models in the CCS. The first of these models couples an ensemble of guiding center trajectories to an MHD bulk. By retaining the moving-dipole contribution to the guiding center magnetization, and adopting the variational guiding center equations of Littlejohn, this model achieves exact energy and momentum balance. In contrast, previous current-coupling drift-kinetic-MHD hybrids conserve neither energy nor momentum. Our second model couples an ensemble of hot gyrocenters to an MHD bulk.   By employing gyrocenter equations of motion derived from a new gauge-invariant single-particle gyrocenter Lagrangian, this model enjoys exact conservation laws for energy, momentum, and hot charge. In contrast, previous current-coupling gyrokinetic-MHD models fail to conserve momentum and hot charge when the background magnetic field is nonuniform. {\color{black}It may be important to point out that while both the drift kinetic and gyrokinetic models require $B/|\nabla\bB|\gg \rho_L$ in order to be physically sound, the models' exact conservation laws persist even when this assumption is violated.}

The importance of incorporating an exact energy balance into hybrid Vlasov-MHD models has been thoroughly examined in \cite{TrTaCaMo}. Those authors have shown that failure to achieve an exact energy balance leads to the presence of unphysical instabilities. Likewise, exact momentum balance is known to be an important ingredient in models used to study intrinsic rotation in tokamaks \cite{Parra_Barnes_2015}, which must accurately track the transfer of angular momentum from kinetic turbulence into rotation of the bulk plasma. We therefore believe that, going forward, our new hybrid  models will play an important role in nonlinear current-coupling hybrid simulations of tokamaks and other fusion devices. {\color{black}In addition, we remark that hybrid CCS theories could also be used to describe kinetic electron populations, thereby extending their range of applicability.}

While we have emphasized the current-coupling approach to hybrids over the pressure-coupling approach in this paper, our results should still be viewed as important to practitioners of the PCS. This is essentially because existing pressure-coupling drift-kinetic and gyrokinetic hybrids appear to suffer from similar issues with conservation laws as their current-coupling cousins. The models presented here, being free of such issues, may be used to form the basis for new conservative pressure-coupling hybrids in the low-frequency approximation. We hope to explore this particular topic in a future publication. {\color{black}A possible strategy in this direction would exploit the Hamiltonian structure underlying the hybrid models presented here; this is a matter of current investigation}.

\medskip

\paragraph{Acknowledgments.} We are greatly indebted to Alain J. Brizard for enlightening certain interpretative questions that emerged in the first version of this work. We also wish to thank Elena Belova, Amitava Bhattacharjee, Alexander R. Close, Greg Hammett, Russell M. Kulsrud, Matthew Kunz, Philip J. Morrison, Hong Qin, Jesus J. Ramos, Paul M. Skerritt, Timothy Stoltzfus-Dueck, Harold Weitzner and Roscoe B. White for their valuable feedback. J.W.B. acknowledges financial support from U.S. Department of Energy, Office of Science, Fusion Energy Sciences under Award No. DE-FG02-86ER53223.
C.T. acknowledges financial support from the Leverhulme Trust Research Project Grant No. 2014-112, and by the London Mathematical Society Grant No. 31439 (Applied Geometric Mechanics Network).

\bigskip

\appendix

\section{Variational structure of collisionless kinetic theories}

This appendix develops the variational theory underlying collisionless kinetic theory and establishes links and connections between the phase-space Lagrangian for particle trajectories and Low's variational principle \cite{Low} for Lagrangian paths on phase-space. Eventually, the reduction from Lagrangian paths to Eulerian trajectories is performed by applying Euler-Poincar\'e theory \cite{HoMaRa1998}. {\color{black}This procedure has rather general validity and has been expolited in many contexts, including both fluid theories \cite{Keramidas} and fully kinetic treatments \cite{CaTr}.}

This appendix contains the material that is used to treat the kinetic component of hybrid MHD models throughout the paper. The fluid MHD part is treated in Eulerian coordinates as in the standard approach by Newcomb \cite{Newcomb} {\color{black}(for a previous variational approach to charged fluid dynamics, see \cite{Katz})}.

\subsection{From phase-space actions to Low's Lagrangian on phase-space}

We begin by writing the well known phase-space action principle for a point particle in an external electromagnetic field:
\beq
\delta \int_{t_1}^{t_2}\!\left[(m{\bf v}+q{\bf A})\cdot\dot{\bf x}-\frac{m}2|\bv|^2-q\varphi\right]\,{\rm d} t=0
\label{PSAction}
\eeq
producing Hamilton's equations
\beq
\dot{\bx}=\bv
\,,\qquad
\dot{\bv}=\frac{q}{m}\left(-\partial_t\bA-\nabla\varphi+\bv\times\bB\right)
\label{hamilton}
\eeq
for the particle trajectory $\bz(t)=(\bx(t),\bv(t))$ on phase-space. In order to introduce Lagrangian coordinates, we fix a specific initial condition $\bz(0)=\boldsymbol\zeta$ and introduce Lagrangian phase-space paths $\boldsymbol{z}(\boldsymbol\zeta,t)$ such that
\beq
\bz(t)=\boldsymbol{z}(\boldsymbol\zeta,t)=\big(\boldsymbol{x}(\boldsymbol\zeta,t),\boldsymbol{v}(\boldsymbol\zeta,t)\big)
\label{paths}
\eeq
and $\bz(0)=\boldsymbol{z}(\boldsymbol\zeta,0)=\boldsymbol\zeta$. At this point, the action principle \eqref{PSAction} can be easily rewritten as
\begin{multline}
\delta \int_{t_1}^{t_2}\!\int\!F_0(\bz_0)\Big[(m{\boldsymbol{v}(\bz_0,t)}+q{\bf A}(\boldsymbol{x}(\bz_0,t)))\cdot\dot{\boldsymbol{x}}(\bz_0,t)
\\
-\frac{m}2|\boldsymbol{v}(\bz_0,t)|^2-q\varphi(\boldsymbol{x}(\bz_0,t))\Big]{\rm d}^6\bz_0\,{\rm d} t=0
\,,
\label{LOW1}
\end{multline}
where we have defined $
F_0(\bz_0)=\delta(\bz_0-\boldsymbol\zeta)$. 
It is clear that, although the variational principle \eqref{LOW1} applies to the Lagrangian phase-space path of a point particle, the same principle applies to any collisionless phase-space ensemble upon smoothening the delta distribution. In this latter case, the equations \eqref{hamilton} for a point particle trajectory are replaced by the following equations for Lagrangian phase-space paths:
\beq
\dot{\boldsymbol{x}}=\boldsymbol{v}
\,,\qquad
\dot{\boldsymbol{v}}=\frac{q}{m}\Big(-\partial_t \bA(\boldsymbol{x})-\nabla\varphi(\boldsymbol{x})+\boldsymbol{v}\times\bB(\boldsymbol{x})\Big)
\,.
\label{hamilton2}
\eeq

Let us now rewrite the variational principle \eqref{LOW1} as
\begin{multline}
\delta \int_{t_1}^{t_2}\!\int\!F_0(\bz_0)\Big[\frac{m}2|\dot{\boldsymbol{x}}(\bz_0,t)|^2+q{\bf A}(\boldsymbol{x}(\bz_0,t)))\cdot\dot{\boldsymbol{x}}(\bz_0,t)-q\varphi(\boldsymbol{x}(\bz_0,t))
\\
-\frac{m}2|\dot{\boldsymbol{x}}(\bz_0,t)-\boldsymbol{v}(\bz_0,t)|^2\Big]{\rm d}^6\bz_0\,{\rm d} t=0
\,.
\label{LOW1-2}
\end{multline}
We notice that if the last term is dropped, one recovers exactly the Low action in \cite{Low} in the case of a static external electromagnetic field. The last term has the role of enforcing the relation $\dot{\boldsymbol{x}}(\bz_0,t)=\boldsymbol{v}(\bz_0,t)$, which is imposed {\it ad hoc} in Low's treatment. This last property was noted already in \cite{CeMaHo}, although the last term in  \eqref{LOW1-2} appears there with a different sign: it can be surprising that the sign in this term is irrelevant to the purpose of obtaining the correct dynamics. The minus sign first appeared in \cite{SqQiTa,Tronci2014}, respectively in the context of gyrokinetic theory and magnetic reconnection models.

For later purpose, we rewrite the action \eqref{LOW1-2} as
\[
\delta \int_{t_1}^{t_2}\!\mathcal{L}_{F_0}(\boldsymbol{z},\dot{\boldsymbol{z}})\,{\rm d} t=0
\]
with the definitions
\[
\mathcal{L}_{F_0}(\boldsymbol{z},\dot{\boldsymbol{z}})=\int\! F_0\,l(\boldsymbol{z},\dot{\boldsymbol{z}})\,{\rm d}^6\bz_0
\,,\qquad\ 
l(\boldsymbol{z},\dot{\boldsymbol{z}})=(m{\boldsymbol{v}}+q{\bf A}(\boldsymbol{x}))\cdot\dot{\boldsymbol{x}} 
-\frac{m}2|\boldsymbol{v}|^2-q\varphi(\boldsymbol{x})
\]
Notice that when the system is coupled to Maxwell's equations, one has
\[
\delta \int_{t_1}^{t_2}\!\mathcal{L}(\boldsymbol{z},\dot{\boldsymbol{z}},\bA,\dot{\bA},\varphi,\dot{\varphi})\,{\rm d} t=0
\]
with the definitions
\begin{align*}
&\mathcal{L}=\mathcal{L}_{F_0}(\boldsymbol{z},\dot{\boldsymbol{z}},\bA,\dot{\bA},\varphi,\dot{\varphi})+\mathcal{L}_{EM}(\bA,\dot{\bA},\varphi,\dot{\varphi})
\\
&\mathcal{L}_{EM}(\bA,\dot{\bA},\varphi,\dot{\varphi})=\frac{\varepsilon_0}2\int\left|-\partial_t\bA-\nabla\varphi\right|^2\,\de^3 x+\frac1{2\mu_0}\int|\nabla\times\bA|^2\,\de^3 x
\\
&\mathcal{L}_{F_0}(\boldsymbol{z},\dot{\boldsymbol{z}},\bA,\dot{\bA},\varphi,\dot{\varphi})=\int\! F_0\,l(\boldsymbol{z},\dot{\boldsymbol{z}},\bA,\dot{\bA},\varphi,\dot{\varphi})\,{\rm d}^6z_0
\\
&l(\boldsymbol{z},\dot{\boldsymbol{z}},\bA,\dot{\bA},\varphi,\dot{\varphi})=(m{\boldsymbol{v}}+q{\bf A}(\boldsymbol{x},t))\cdot\dot{\boldsymbol{x}} 
-\frac{m}2|\boldsymbol{v}|^2-q\varphi(\boldsymbol{x},t)
\end{align*}
so that Gauss and Ampere's laws arise as usual from the variational Euler-Lagrange equations
\[
\frac{\partial}{\partial t}\frac{\delta \mathcal{L}}{\delta\dot\varphi}-\frac{\delta \mathcal{L}}{\delta\varphi}=0
\,,\qquad
\frac{\partial}{\partial t}\frac{\delta \mathcal{L}}{\delta\dot{\bA}}-\frac{\delta \mathcal{L}}{\delta\bA}=0
\,.
\]
Here, the functional derivatives are defined as usual: let $F(f)$ be a functional of a function $f(\bx)$ in physical space, then
\[
\delta F=\int\!\frac{\delta F}{\delta f}\,\delta f\,{\rm d}^3 x
\,.
\]
This definition is easily extended to  functionals of several functions in different spaces. For example, the equations of motion \eqref{hamilton2} arise from the Euler-Lagrange equations
\[
\frac{\partial}{\partial t}\frac{\delta \mathcal{L}}{\delta\dot{\boldsymbol{z}}}-\frac{\delta \mathcal{L}}{\delta{\boldsymbol{z}}}=0
\,.
\]

\subsection{Euler-Poincar\'e reduction and Eulerian variational principles\label{reduction_and_eulerian}}

In the previous Section, we considered only Lagrangian paths on phase-space $\boldsymbol{z}(\z_0,t)$. However, collisionless kinetic theories are typically written in terms of the evolution equation for the Eulerian phase-space density $F(\bz,t)$. The latter is defined by the relation $F(\boldsymbol{z},t)\,{\rm d}^6 \boldsymbol{z}=F_0(\bz_0)\, {\rm d}^6 {\bz}_0$, which is equivalent to the following Lagrange-to-Euler map
\[
F(\bz,t)=\int \!F_0(\bz_0)\,\delta\big(\bz-\boldsymbol{z}(\bz_0,t)\big)\, {\rm d}^6 {\bz}_0
\,.
\]
In this Section, we shall perform the reduction from Lagrangian to Eulerian variables by following Euler-Poincar\'e reduction \cite{CeMaHo,HoMaRa1998,SqQiTa}. 
Upon introducing the phase-space vector field $\boldsymbol{\mathcal{X}}$ such that $\dot{\boldsymbol{z}}=\boldsymbol{\mathcal{X}}(\boldsymbol{z})$, one easily rewrites the action \eqref{LOW1} in Eulerian form as
\begin{equation}
\delta \int_{t_1}^{t_2}\!\int\!F(\bz,t)\Big[(m{{\bv}}+q{\bf A}({\bx}))\cdot\mathbf{u}(\bz,t)
-\frac{m}2|{\bv}|^2-q\varphi({\bx})\Big]{\rm d}^6{\bz}\,{\rm d} t=0
\,,
\label{EP1}
\end{equation}
where $\mathbf{u}(\bz,t)$ is the spatial component of the vector field $\boldsymbol{\mathcal{X}}(\bz,t)=(\mathbf{u}(\bz,t),\mathbf{a}(\bz,t))$. Explicitly, one has $(\dot{\boldsymbol{x}},\dot{\boldsymbol{v}})=(\mathbf{u}(\boldsymbol{z},t),\mathbf{a}(\boldsymbol{z},t))$. At this stage we introduce the notation
\beq
\ell(\boldsymbol{\mathcal{X}})=(m{{\bv}}+q{\bf A}({\bx}))\cdot\mathbf{u}(\bz,t)
-\frac{m}2|{\bv}|^2-q\varphi({\bx})
\label{ellsmall}
\eeq
so that the Euler-Poincar\'e variational principle \eqref{EP1} may be rewritten as
\[
\delta \int_{t_1}^{t_2}\! \mathscr{L}(\boldsymbol{\mathcal{X}},F)
\,{\rm d} t=
\delta \int_{t_1}^{t_2}\!\int\!F \,\ell(\boldsymbol{\mathcal{X}})\,
{\rm d}^6{\rm z}\,{\rm d} t=0
\,.
\]
The Euler-Poincar\'e reduction is based on the following symmetry property:
\[
\mathcal{L}_{F_0}(\boldsymbol{z},\dot{\boldsymbol{z}})=\int\!F \,\ell(\boldsymbol{\mathcal{X}})\,
{\rm d}^6{\rm z}=\mathscr{L}(\boldsymbol{\mathcal{X}},F)
\,.
\]
The problem is now to find the explicit expression of the Eulerian variations $\delta\boldsymbol{\mathcal{X}}$ and $\delta F$. On one hand, the definition of $F$ (e.g. the Lagrange-to-Euler map) enforces
\[
\delta F = -\nabla_{\bz}\cdot( F \boldsymbol\Xi)
\,,
\]
where $\boldsymbol\Xi$ is such that $\delta \boldsymbol{z}=\boldsymbol\Xi(\boldsymbol{z},t)$. 
On the other hand, the variation $\delta\boldsymbol{\mathcal{X}}$ must be computed by using the relation $\boldsymbol{\mathcal{X}}(\bz,t)=\dot{\boldsymbol{z}}(\boldsymbol{z}^{-1}(\bz,t),t)$, where $\boldsymbol{z}^{-1}$ denotes the inverse of the Lagrangian path $\boldsymbol{z}$. Vector calculus calculations and extensive use of the properties of Jacobian matrices lead to the relation
\[
\delta\boldsymbol{\mathcal{X}}=\partial_t\boldsymbol\Xi+(\boldsymbol{\mathcal{X}}\cdot\nabla_\z)\boldsymbol{\Xi}-(\boldsymbol\Xi\cdot\nabla_\z)\boldsymbol{\mathcal{X}}
\,.
\]
These relations yield the following Euler-Poincar\'e equations for an arbitrary functional $\mathscr{L}(\boldsymbol{\mathcal{X}},F)$:
\beq
\frac{\partial}{\partial t}\frac{\delta \mathscr{L}}{\delta\boldsymbol{\mathcal{X}}} +
(\boldsymbol{\mathcal{X}}\cdot\nabla_\z)\frac{\delta \mathscr{L}}{\delta\boldsymbol{\mathcal{X}}}+\nabla_\z\boldsymbol{\mathcal{X}}\cdot \frac{\delta \mathscr{L}}{\delta\boldsymbol{\mathcal{X}}}=F\,\nabla_\z\frac{\delta \mathscr{L}}{\delta F}
\,,
\label{EPEQ1}
\eeq
while the kinetic transport equation arises by taking the time-derivative of the Lagrange-to-Euler map, that is
\beq
\partial_t F+\nabla_{\bz}\cdot( F \boldsymbol{\mathcal{X}})=0\,.
\label{vlasov1}
\eeq
The standard Vlasov equation arises by replacing
\[
\frac{\delta \ell}{\delta\boldsymbol{\mathcal{X}}}=\left(\frac{\delta \ell}{\delta\mathbf{u}},\frac{\delta \ell}{\delta\mathbf{a}}\right)=\bigg(m\bv+q\bA(\bx),\,0\bigg)
\]
into \eqref{EPEQ1} thereby obtaining
\[
\boldsymbol{\mathcal{X}}(\bz,t)=\big(\mathbf{u}(\bz,t),\mathbf{a}(\bz,t)\big)=\left(\bv,\,\frac{q_h}{m_h}\big(-\partial_t\bA-\nabla\varphi(\bx)+\bv\times\bB(\bx)\big)\right)
\]
Replacing this in \eqref{vlasov1} yields the usual Vlasov equation
\[
\partial_t F+\bv\cdot\nabla F+
\frac{q_h}{m_h}\big(-\partial_t\bA-\nabla\varphi(\bx)+\bv\times\bB(\bx)\big)\cdot\nabla_{\bv}F=0
\]

Notice that, when the Vlasov kinetic equation is coupled to Maxwell's equations, the functional \eqref{ellsmall} becomes of the type $\ell(\boldsymbol{\mathcal{X}},\bA,\dot{\bA},\varphi,\dot{\varphi})$ (analogously, $\mathscr{L}(\boldsymbol{\mathcal{X}},F)$ is replaced by $\mathscr{L}(\boldsymbol{\mathcal{X}},F,\bA,\dot{\bA},\varphi,\dot{\varphi})$) and the action principle \eqref{EP1} becomes
\begin{equation}
\delta\! \int_{t_1}^{t_2}\!\left(\mathscr{L}+\mathcal{L}_{EM}\right){\rm d} t=
\delta \!\int_{t_1}^{t_2}\!\left[\int\!F\ell(\boldsymbol{\mathcal{X}},\bA,\dot{\bA},\varphi,\dot{\varphi})\,{\rm d}^6{\rm z}+\mathcal{L}_{EM}(\bA,\dot{\bA},\varphi,\dot{\varphi})\right]{\rm d} t=0
\,.
\label{EP2}
\end{equation}
 Then, Gauss' and Amp\`ere's Law arise from the Euler-Lagrange equations
\[
\frac{\partial}{\partial t}\frac{\delta {L}}{\delta\dot\varphi}-\frac{\delta {L}}{\delta\varphi}=0
\,,\qquad
\frac{\partial}{\partial t}\frac{\delta {L}}{\delta\dot{\bA}}-\frac{\delta {L}}{\delta\bA}=0
\,,
\]
where $L(\boldsymbol{\mathcal{X}},F,\bA,\dot{\bA},\varphi,\dot{\varphi})=\mathscr{L}(\boldsymbol{\mathcal{X}},F,\bA,\dot{\bA},\varphi,\dot{\varphi})+\mathcal{L}_{EM}(\bA,\dot{\bA},\varphi,\dot{\varphi})$.

\subsection{Extension to drift-kinetic and gyrokinetic theories\label{sec:appGC}}
The previous construction does not involve substantial modifications when it is adapted to drift-kinetic and gyrokinetic theories. In the first case, a full treatment was recently given in \cite{BrTr}. The six dimensional phase-space is replaced by a 4-dimensional space with coordinates $\z_{\rm gc}=(\bX,v_\|)$ and the phase-space action principle \eqref{PSAction} is adapted to a single guiding center trajectory  as follows \cite{Littlejohn}:
\beq\label{VarPrincGC-sP}
\delta\!\int_{t_1}^{t_2}\!\left[\Big(mv_\|\bb(\bX,t)+q\bA(\bX,t)\Big)\cdot\dot{\mathbf{X}}-\frac{m}2v_\|^2-\mu B(\bX)-q_h\varphi(\bX)\right]\,\de t=0\,,
\eeq
where we have introduced the notation $B=|\bB|$ and $\bb=\bB/B$. The guiding center motion arises from the Euler-Lagrange equations, thereby yielding 
\beq\label{SP-EQ}
\dot{\bX}=\frac1 {B^*_\|}\Big(v_\|\bB^*-\bb\times\bE^*\Big)
\,,\qquad \ 
\dot{v}_\|= \frac{q}{mB^*_\|}\,\bB^*\cdot\bE^*,
\eeq
where
\beq
\bB^*:=\nabla\times\left(\bA+a_h^{-1}v_\parallel \bb\right)
\,,\qquad\ 
\bE^*:=-\partial_t\left(\bA+a_h^{-1}v_\parallel \bb\right)-\nabla(\varphi+\mu B)
\label{EffFields2}
\eeq
and all field variables in \eqref{SP-EQ} are evaluated at the guiding center position $\bX$. Notice that the magnetic moment $\mu$ appears in the theory as a constant parameter. This implies that, as noticed in \cite{BrTr}, Lagrangian paths will also be parameterized by the magnetic moment, so that the analogue of \eqref{paths} reads
\beq
\bz_{\rm gc}(t;\mu)=\boldsymbol{z}_{\rm gc}(\boldsymbol\zeta,t;\mu)=\big(\boldsymbol{X}(\boldsymbol\zeta,t;\mu),\mathcal{V}_\parallel(\boldsymbol\zeta,t;\mu)\big)
\,.
\label{paths2}
\eeq
The rest of the construction can be reproduced by following exactly the same steps as in the previous Sections. 

Similar arguments also apply to gyorkinetic theory, as already shown in \cite{SqQiTa}. The variational principle for a single gyrocenter is given by
\begin{align}
\delta\int_{t_1}^{t_2}\!\ell_{\text{gy}}(\bz,\dot{\bz};\tilde{\varphi},\tilde{\bA},\tilde{\bE},\tilde{\bB})\,\de t=0,
\end{align} 
where $\ell_{\text{gy}}$ is given in Eq.\,(\ref{new_gy_lag}) and $\bz=\bz(t)$ is the trajectory of a single gyrocenter. The Euler-Lagrange equations associated with this variational principle are
\begin{gather}
q_h\bE^*-q_h\bB^*\times\dot{\bX}-m_h \dot{v}_{\parallel}\bb_{\text{eq}}=0\\
m_h\dot{\bX}\cdot\bb_{\text{eq}}-m_h v_\parallel=0,
\end{gather}
where
\begin{align}
\bB^*&=\bB(\bX)+m_hq_h^{-1}v_\parallel\nabla\times\bb_{\text{eq}}+\nabla\times\langle\!\langle\tilde{\bB}(\bX+\lambda\brho)\times\brho\rangle\!\rangle\\
\bE^*&=\tilde{\bE}(\bX)-q_h^{-1}\nabla([\mu+\delta\mu]B_{\text{eq}})+\langle\!\langle(\nabla\times\tilde{\bE})(\bX+\lambda\brho)\times\brho\rangle\!\rangle\nonumber\\
&+\nabla\langle\!\langle\tilde{\bE}(\bX+\lambda\brho)\cdot\brho\rangle\!\rangle.
\end{align}
The corresponding Eulerian variational principle for a distribution of gyrocenters in prescribed electromagnetic fields is therefore given by
\begin{align}
\delta\int_{t_1}^{t_2}\iint_\mu\ell_{\text{gy}}(\bz,\bcalX_{\text{gy}}(\bz);\tilde{\varphi},\tilde{\bA},\tilde{\bE},\tilde{\bB})F\,\,dv_\parallel \de t=0,
\end{align}
where $\bcalX_{\text{gy}}=(\mathbf{u}_{\text{gy}},a_{\parallel\text{gy}})$ is the Eulerian phase-space fluid velocity and the appropriate variational constraints are given in \ref{reduction_and_eulerian}. The Euler-Lagrange equations are now
\begin{gather}
q_h\bE^*-q_h\bB^*\times\mathbf{u}_{\text{gy}}-m_h a_{\parallel\text{gy}}\bb_{\text{eq}}=0\\
m_h\mathbf{u}_{\text{gy}}\cdot\bb_{\text{eq}}-m_h v_\parallel=0,
\end{gather}
where $\bX$ and $v_\parallel$ are Eulerian phase space coordinates instead of the coordinates of a single gyrocenter trajectory.

\end{document}